\documentclass[12pt]{article}

\usepackage{graphicx,psfrag,epsf}
\usepackage{natbib}
\usepackage{mathtools}
\usepackage[english]{babel} % English language/hyphenation
\usepackage{microtype} % Better typography
\usepackage{amsmath,amsfonts,amsthm}
\usepackage{chngcntr}
\usepackage[toc,page]{appendix}
\usepackage{ amssymb }
\usepackage{array}
\usepackage{natbib}
\usepackage{setspace}
\usepackage{fix-cm}
\usepackage{url}
\usepackage{tikz}
\usepackage{arydshln}
\usepackage{cool}
\usepackage{enumitem}
\usepackage{soul}
\usepackage{threeparttable}
\usepackage{rotating}
\usepackage{tablefootnote}
\usepackage{color}
\usepackage{tabularx}
\usepackage[font = small,labelfont=bf,textfont=it]{caption} % Custom captions under/above floats in tables or figures
\usepackage{footnote}
\usepackage{algorithm}% http://ctan.org/pkg/algorithms
\usepackage[noend]{algpseudocode}% http://ctan.org/pkg/algorithmicx
\usepackage{footnote}
\usepackage{multirow, makecell}
\usepackage{diagbox}
\makeatletter
\newcommand{\distas}[1]{\mathbin{\overset{#1}{\kern\z@\sim}}}%
\newcommand{\bm}[1]{\mathbf{#1}}
\newsavebox{\mybox}\newsavebox{\mysim}
\newcommand{\distras}[1]{%
  \savebox{\mybox}{\hbox{\kern3pt$\scriptstyle#1$\kern3pt}}%
  \savebox{\mysim}{\hbox{$\sim$}}%
  \mathbin{\overset{#1}{\kern\z@\resizebox{\wd\mybox}{\ht\mysim}{$\sim$}}}%
}

\newtheorem{proposition}{Proposition}

\newcolumntype{C}[1]{>{\centering\let\newline\\\arraybackslash\hspace{0pt}}m{#1}}

\newcommand{\be}{\begin{equation}}
\newcommand{\ee}{\end{equation}}
\newcommand{\bi}{\begin{itemize}}
\newcommand{\ei}{\end{itemize}}
\newcommand{\ben}{\begin{enumerate}}
\newcommand{\een}{\end{enumerate}}
\newcommand{\stb}{\State $\bullet$ \;}
\newcommand{\crd}[1]{{\color{red}{#1}}}
\newcommand{\cbl}[1]{{\color{blue}{#1}}}
\newcommand{\crdb}[1]{{\color{red}{\textbf{#1}}}}
\newcommand{\cblb}[1]{{\color{blue}{\textbf{#1}}}}
\newcommand{\crdbm}[1]{{\color{red}{\mathbf{#1}}}}
\newcommand{\cblbm}[1]{{\color{blue}{\mathbf{#1}}}}
\allowdisplaybreaks
\DeclareMathOperator*{\argmin}{\arg\!\min}

\makeatletter
\def\toprule{%
  \noalign{\ifnum0=`}\fi\hrule \@height \thickarrayrulewidth \futurelet
   \reserved@a\@xthickhline}
\def\@xthickhline{\ifx\reserved@a\thickhline
               \vskip\doublerulesep
               \vskip-\thickarrayrulewidth
             \fi
      \ifnum0=`{\fi}}
\makeatother

\newlength{\thickarrayrulewidth}
\makeatother

\let\oldbibliography\thebibliography
\renewcommand{\thebibliography}[1]{\oldbibliography{#1}
\setlength{\itemsep}{0pt}} %Reducing spacing in the bibliography.

\newcommand{\blind}{1}

\patchcmd{\footnotemark}{\stepcounter{footnote}}{\refstepcounter{footnote}}{}{}
\newcounter{savecntr}% Save footnote counter
\newcounter{restorecntr}% Restore footnote counter
\begin{document}

\def\spacingset#1{\renewcommand{\baselinestretch}%
{#1}\small\normalsize} \spacingset{1}

\if1\blind
{
   \title{Analysis-of-marginal-Tail-Means (ATM): a robust method for discrete black-box optimization}
  \author{Simon Mak\setcounter{savecntr}{\value{footnote}}\thanks{School of Industrial and Systems Engineering, Georgia Institute of Technology}\; $^{\ddag}$, \; C. F. Jeff Wu\setcounter{restorecntr}{\value{footnote}}%
  \setcounter{footnote}{\value{savecntr}}\footnotemark% Print footnotemark
  \setcounter{footnote}{\value{restorecntr}}
\footnote{Corresponding author} 
\footnote{This work is supported by the U. S. Army Research Office under grant number W911NF-17-1-0007.}
}
  \maketitle
} \fi

\if0\blind
{
   \title{Analysis-of-marginal-Tail-Means (ATM): a robust method for discrete black-box optimization}
   \date{}
   \maketitle
} \fi

\bigskip

\begin{abstract}

\noindent We present a new method, called Analysis-of-marginal-Tail-Means (ATM), for effective robust optimization of discrete black-box problems. ATM has important applications to many real-world engineering problems (e.g., manufacturing optimization, product design, molecular engineering), where the objective to optimize is black-box and expensive, and the design space is inherently discrete. One weakness of existing methods is that they are not robust: these methods perform well under certain assumptions, but yield poor results when such assumptions (which are difficult to verify in black-box problems) are violated. ATM addresses this via the use of marginal tail means for optimization, which combines both rank-based and model-based methods. The trade-off between rank- and model-based optimization is tuned by first identifying important main effects and interactions, then finding a good compromise which best exploits additive structure. By adaptively tuning this trade-off from data, ATM provides improved robust optimization over existing methods, particularly in problems with (i) a large number of factors, (ii) unordered factors, or (iii) experimental noise. We demonstrate the effectiveness of ATM in simulations and in two real-world engineering problems: the first on robust parameter design of a circular piston, and the second on product family design of a thermistor network.

\end{abstract}

\noindent%
\vfill

\newpage
\spacingset{1.45} % DON'T change the spacing!

\section{Introduction}
\label{sec:intro}

For many real-world engineering problems, e.g., in manufacturing, product design, and molecular engineering, one requires the optimization of some objective function over a large, \textit{discrete} factorial space. This problem is challenging for two reasons. First, the objective function is typically ``black-box'', in that little is known on it prior to experiments. Second, each evaluation of this objective can be expensive, requiring costly physical experiments or simulations. Because of this, only limited evaluations can be used for optimization. For such problems, one weakness of existing methods is that they are not \textit{robust}: these methods perform well under certain modeling assumptions, but yield poor results when such assumptions (difficult to verify in black-box problems) are violated. We propose a new method, called \textit{Analysis-of-marginal-Tail-Means} (ATM), which provides effective and \textit{robust} optimization for a broad range of discrete black-box problems.

%Second, in system-level problems, there are often many decision variables to optimize over, which requires searching over a high-dimensional discrete space.

%A key goal in engineering design is to find parameter settings which optimize some performance metric $f$ over a large, discrete design space. Here, $f$ can measure the quality of a final product, the productivity of a process, or the robustness of a system in the presence of noise factors. There are two challenges implicit in such a parameter optimization problem. First, the response surface of $f$ over input parameters is typically ``black-box'', in that an experimenter has little-to-no prior knowledge on the functional form or properties of $f$. Second, each evaluation of $f$ is computationally or monetarily expensive, so only a limited number of function observations can be used for solving the desired parameter optimization problem. To this end, we propose in this paper a new method called \textit{Analysis-of-marginal-Tail-Means} (ATM), which offers \textit{effective} and \textit{robust} optimization performance for both smooth and rugged response surfaces, using a limited number of function evaluations.

ATM has important applications to many real-world engineering problems, where the design space is inherently discrete. One application is in manufacturing optimization, where the goal is to find an optimal setting of tools, parts and conditions which maximizes product quality. Because of manufacturing complexities and the discrete nature of factors (e.g., parts come in select varieties from a supplier), most manufacturing optimization problems are black-box and combinatorial \citep{DZ2000}. Here, ATM can be used to find a good operation setting with a limited number of expensive experiments. Another application is to the emerging problem of product family design \citep{Nea2002}. The idea is to develop a family of products (e.g., circuit networks) with customizable parts (e.g., resistors), then employ different combinations of parts to target specific market needs. Here, ATM can be used to find a parts specification which caters to each market's needs, using a small number of experiments. Our method is also useful in many molecular and material engineering problems, e.g., designing molecules in pharmaceutical drugs \citep{Mea2006}, or optimizing nanowire growth \citep{Xea2009}.

%, where the goal is to find control settings which jointly maximize product quality and enjoy robustness to noise factors (see \citealp{Tag1986, WH2009} for details). For example, in designing an electric circuit, an engineer may be interested in finding an optimal configuration of capacitors and resistors, which (a) maximizes the power of such a circuit, and (b) minimizes its performance variability in the presence of uncontrollable environmental noise. Our method also finds important applications in the area of combinatorial chemistry in the pharmaceutical industry \citep{Mea2006}, where the aim is to design a molecular compound (from potentially millions of candidate compounds) which is most effective at treating a particular disease. All of these applications encounter the same two challenges mentioned earlier: the metric of interest $f$ is typically black-box, and each function evaluation can be quite expensive to generate. The proposed method offers a novel approach for jointly tackling these two challenges, and can provide more effective and robust optimization performance over existing techniques for the above applications.

Viewed as a parameter optimization problem, there is a notable body of literature which applies to the discrete, black-box problem at hand. An early technique, popularized by \cite{Tag1986}, is the \textit{Analysis-of-marginal-Means} (AM) method, which uses the estimated marginal means of each factor to predict its optimal level setting. While AM enjoys excellent performance with limited samples if the black-box function is nearly additive (this is formalized later in the paper), \cite{Wea1990} showed that AM can yield poor performance when this additivity is violated. To address this, these authors instead advocated for a simple method called \textit{Pick-the-Winner} (PW), which chooses from the \textit{observed} settings the one yielding the best performance (the ``winner''). In the same paper, \cite{Wea1990} introduced a batch-sequential scheme for AM and PW, called \textit{Sequential-Elimination-of-Levels} (SEL), which iteratively eliminates levels with the worst-performing marginal mean or minimum. SEL was further developed by \cite{Mea2006} and \cite{Mea2009}, who employed genetic algorithms and Gaussian processes to guide the elimination procedure.

Methods for \textit{continuous} black-box optimization can also be used, with appropriate discretization, for the discrete problem here. This includes the popular Expected Improvement method (EI; \citealp{Jea1998}) for black-box global optimization, the sequential design strategies in \cite{Wea2000} and \cite{Rea2008}, as well as the minimum energy designs in \cite{Jea2015}. However, the effectiveness of these ``discretized'' methods depends on the type of discrete factors at hand. For \textit{ordinal} discrete factors (i.e., with ordered levels), such methods may perform well due to its inherent continuous nature. For \textit{nominal} discrete factors (i.e., with unordered levels), such methods may perform poorly by accounting for spurious order between levels. Nominal discrete factors, however, are present in many engineering problems, particularly in product family design (which part to use out of four choices?) and molecular engineering (which reagent to add at a molecule position?). To contrast, the AM and PW methods are applicable for both ordinal and nominal factors.

Despite this body of work, a key difficulty noted by practitioners (see \citealp{Sim2004, LP2011}) is that there is no effective \textit{robust} optimization strategy -- some methods perform well under certain conditions, but yield poor results when these conditions are violated. This is a serious problem, because one does not know if such conditions hold for black-box problems prior to experiments. This lack of robustness is nicely illustrated by the AM and PW methods. When the objective to optimize is nearly additive, AM implicitly makes use of a fitted additive model to provide effective optimization with few samples (since marginal means can be well estimated from limited data). However, when strong interactions are present, AM can return wildly suboptimal solutions due to the poor fit of an additive model, and the rank-based PW strategy can yield better performance. In light of this, \cite{Wea1987} suggests that ``optimization should be based on empirical models \textit{as well as} the ranking of ... experimental data''. The proposed ATM method does precisely this, by first (i) removing from each level the worst-performing runs above a threshold (ranking), then (ii) using the marginal means of remaining runs, i.e., the tail means, to predict the optimal setting (modeling). This provides a trade-off between the model-based AM and the rank-based PW, and allows one to exploit \textit{local} marginal structure for optimization. By adaptively tuning this trade-off from data, ATM can provide effective optimization with limited data in a wide range of black-box engineering problems.

This paper is organized as follows. Section 2 outlines the discrete optimization problem at hand, and reviews AM and PW. Section 3 introduces the proposed ATM method, and describes two trade-offs contributing to its robustness in optimization. Section 4 presents a tuning procedure for ATM, as well as a level elimination scheme \texttt{sel.atm} for batch-sequential optimization. Sections 5 and 6 investigate the performance of \texttt{sel.atm} in simulations and in two real-world applications. Section 7 concludes with directions for future research.

%\bi
%\item We reveal some novel insights on the performance trade-off between AM and PW, and when each method does best or worst. In particular, AM and PW each does well in certain situations, but very poorly in others. This is not reassuring to a practictioner, who typically doesn't know a priori what the function properties are, and want a method which performs well under all situations.
%\item ROBUSTNESS: We then propose a unified methodology, called analysis-of-marginal-Tail-Means (ATM), which provides a continuum of predictors with PW prediction on one end and AM prediction on the other. We show that ATM exploits the trade-off between model-based and rank-based optimization to provide \textit{effective} optimization over a \textit{broad} class of response surfaces. 
%\item We then combine ATM with SEL to provide a new sequential optimization procedure called SEL-ATM. Compared to the usual SEL with AM or PW, we show the proposed approach provides more robust optimization performance -- in that it provides excellent performance for both smooth and rough response surfaces. Can also provide better performance by eliminating the correct levels.
%\ei
%\item Roadmap of paper
%\ei

\section{Problem set-up and motivation}
We first introduce the discrete black-box problem at hand, then show via a motivating example why AM and PW are not effective robust optimization methods.

\subsection{Discrete black-box optimization}
Suppose there are $p$ discrete factors of interest (these can be either ordinal or nominal), with the $l$-th factor having $N_l$ levels, $l=1, \cdots, p$. Denote these levels as $[N_l] := \{1, \cdots, N_l\}$. Let $\mathcal{X} = [N_1] \times [N_2] \times \cdots \times [N_p]$ be the feasible space of all level combinations, and let $f:\mathcal{X} \rightarrow \mathbb{R}$ be the objective function to be optimized. Without loss of generality, we assume minimization in this paper. The desired discrete optimization problem becomes:
\begin{equation}
\bm{x}^* := \underset{\bm{x} \in \mathcal{X}}{\text{argmin}}\; f(\bm{x}).
\label{eq:opt}
\end{equation}
The goal is to find the optimal setting $\bm{x}^* \in \mathcal{X}$ using a small number $n$ of (possibly noisy) evaluations $\bm{y} \in \mathbb{R}^n$ on objective $f$. We emphasize again that the feasible space $\mathcal{X}$ should be inherently discrete from the engineering problem at hand.

There are several practical challenges in solving \eqref{eq:opt}. First, the objective $f$ is typically black-box, in that an experimenter has little knowledge on its functional form or properties prior to data. Second, each evaluation of $f$ can be expensive, requiring costly physical experiments or time-consuming simulations. For example, the evaluation of combustion instability in a rocket engine can take several weeks via simulation \citep{Mea2017}, and is extremely costly via physical experiments \citep{Yea2018}. To this end, the \textit{robustness} of an optimization method is paramount -- a good method should learn structure on $f$ from data, and exploit this structure for effective optimization with limited objective evaluations.

Below, we review the AM and PW methods (which are purely model- and rank-based methods, respectively), and show via two examples why these methods do not enjoy the desired robustness. This motivates the proposed ATM technique, which employs an adaptive trade-off between model- and rank-based methods to achieve effective robust optimization.

\subsection{Analysis-of-marginal-Means and Pick-the-Winner}
\label{sec:ampw}

\begin{figure}[t]
\begin{minipage}{0.42\textwidth}
\centering
\includegraphics[width=0.95\textwidth]{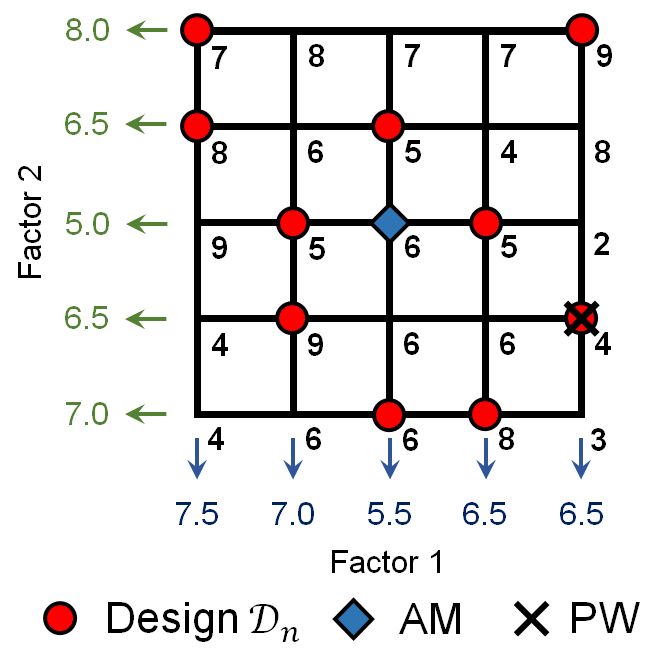}
\captionof{figure}{A visualization of AM and PW for a 2-d toy example. Blue and green arrows show estimated marginal means for factors 1 and 2.}
\label{fig:ampw}
\end{minipage}
\hspace{0.2cm}
\begin{minipage}{0.57\textwidth}
\small
\centering
\begin{tabular}{c c  c  c  c }
\toprule
 \multicolumn{5}{c}{\textbf{Friedman}}\\
% \hline
Method & $n=25$ & $n=50$ & $n=75$ & $n = \#\{\mathcal{X}\} = 5^5$\\
\toprule
AM & \crdb{\textbf{2.13}} & \crdb{\textbf{1.95}} & \crdb{\textbf{1.87}} & \crdb{\textbf{1.81}} \\
PW & 5.33 & 4.21 & 4.14 & \crdb{\textbf{1.81}} \\
\toprule
\multicolumn{5}{c}{\textbf{DetPep10}} \\
%\hline
Method & $n=25$ & $n=50$ & \multicolumn{2}{c}{ $n = \#\{\mathcal{X}\} = 5^3$}\\
\toprule
AM & 11.45 & 11.42 & \multicolumn{2}{c}{11.41} \\
PW & \crdb{\textbf{11.22}} & \crdb{\textbf{10.98}} & \multicolumn{2}{c}{\crdb{\textbf{10.83}}} \\
\toprule
\end{tabular}
\normalsize
\captionof{table}{Mean predicted minimum $f(\hat{\bm{x}})$ using AM and PW, averaged over 100 random OAs of run size $n$. The best method is colored \crd{red}.}
\label{tbl:toill}
\end{minipage}
\end{figure}

Analysis-of-marginal-Means (AM), popularized by \cite{Tag1986}, is a well known parameter optimization method for solving \eqref{eq:opt}. Given the constraint of limited data, the key idea in AM is to exploit the marginal means of each factor (which can be well estimated from, say, an orthogonal array (OA)) for guiding optimization. More specifically, let $m_l(x_l)$ be the true marginal mean of factor $l$ at level $x_l \in [N_l]$, and let $\hat{m}_l(x_l)$ be its estimated marginal mean from observed data. The AM predictor of the optimal setting $\bm{x}^*$ can be written as:
\begin{equation}
\hat{\bm{x}}_{\rm AM} = (\hat{x}_1, \cdots, \hat{x}_p), \quad \hat{x}_l := \underset{x_l \in [N_l]}{\text{argmin}} \; \hat{m}_l(x_l), \quad l = 1, \cdots, p.
\label{eq:am}
\end{equation}
In words, AM selects for each factor $l$ the level $\hat{x}_l$ with lowest estimated marginal mean, then uses the level combination $\hat{\bm{x}}_{\rm AM} = (\hat{x}_1, \cdots, \hat{x}_p)$ to predict $\bm{x}^*$.

Figure \ref{fig:ampw} visualizes AM using a 2-d toy example with $N_l = 5$ levels for each factor (for simplicity, assume $f$ is observed without noise). Here, the black grid represents the feasible settings in $\mathcal{X}$ (with function values in black), and the red dots mark the observed settings. The blue and green arrows show the estimated marginal means (computed with two data points) for the two factors, which approximate the true marginal means (computed with five points). For both factors, the estimated marginal means are minimized at level 3, so the AM predictor of the optimal setting $\bm{x}^*$ is $\hat{\bm{x}}_{\rm AM} = (3,3)$, with objective value $f(\hat{\bm{x}}_{\rm AM})=6$.

The advantage of AM is that, whenever the objective $f$ follows the additive model:
\begin{equation}
f(\bm{x}) = \sum_{l=1}^p f_l(x_l),
\label{eq:add}
\end{equation}
AM exploits the underlying marginal structure to provide effective optimization with a \textit{small} number of evaluations. To see why, suppose $f$ follows the additive, main-effect model in \eqref{eq:add}. If the true marginal means $m_l(x_l)$ are known for all factors $l = 1, \cdots, p$ and all levels $x_l \in [N_l]$, then it is easy to see that the AM predictor \eqref{eq:am} (with $m_l(x_l)$ replacing $\hat{m}_l(x_l)$) returns the optimal setting $\bm{x}^*$. These marginal means can also be reliably estimated with a small number of runs from a balanced experimental design, since each data point is used multiple times in the AM predictor $\hat{\bm{x}}_{\rm AM}$. For the toy example in Figure \ref{fig:ampw}, each data point (in red) is used twice in $\hat{\bm{x}}_{\rm AM}$: once for estimating the marginal means of the first factor, and another time for the second factor. This ``reuse'' of data allows AM to \textit{augment} the effective sample size on $f$, thereby providing effective optimization with few evaluations.

However, AM is effective \textit{only} when $f$ is additive (or nearly additive, see \citealp{Wea1987} and Proposition 1 later). When a moderately large interaction is present, AM can return poor solutions which can be worse than observed settings! To address this concern, \cite{Wea1990} advocated for a Pick-the-Winner (PW) approach, which simply chooses the best \textit{observed} setting as the optimal solution. Figure \ref{fig:ampw} visualizes PW for the 2-d toy example. Assuming again no noise, PW selects the point with the lowest observed value, which here is $\hat{\bm{x}}_{\rm PW} = (5,4)$ with objective value $4$. For this example, the simple PW approach outperforms the AM method, which returns an objective value of 6. This suggests the presence of strong interactions in $f$, which is indeed true from an inspection of function values in Figure \ref{fig:ampw}.

\subsection{Two motivating examples}
\label{sec:ampwex}

%The above marginal optimization framework includes the Analysis-of-marginal-Means method \citep{Tag1986} and the Pick-the-Winner method \citep{Wea1990} as special instances. To see this, set the marginal statistic $\mathcal{M}_l$ as the marginal mean $\mathcal{M}_{\rm mean}$ in \eqref{eq:mmean} for all factors $l = 1, \cdots, p$. The marginal predictor $\hat{\bm{x}}$ then selects, for each factor, the level which minimizes the estimated marginal means. This is precisely the AM method popularized by \cite{Tag1986}, which employs marginal mean estimates to choose the optimal level setting for each factor. Likewise, setting $\mathcal{M}_l$ as the marginal minimum $\mathcal{M}_{\rm min}$ in \eqref{eq:mmin}, the marginal predictor $\hat{\bm{x}}$ selects, for each factor, the level with the smallest observed function value. However, because the smallest observed value over all levels is the smallest observed value in the design, $\hat{\bm{x}}$ picks the setting corresponding to the smallest observed function evaluation. This then reduces to the PW method advocated by \cite{Wea1990}.

We now show by using two examples why both AM and PW are not effective robust optimization methods, in that both can yield poor solutions to \eqref{eq:opt} in certain situations. Consider the two test functions -- the 5-d Friedman function \citep{Fea1983}:
\vspace{-0.2cm}
\begin{equation}
f(\bm{x}) = 10 \sin (\pi x_1 x_2) + 20(x_3 - 0.5)^2 + 10x_4 + 5x_5, \quad \bm{x} \in [0,1]^5,
\label{eq:friedman}
\end{equation}
\vspace{-0.2cm}
and the 3-d DetPep10 function:
\vspace{-0.2cm}
\begin{equation}
f(\bm{x}) = 4(x_1- 2+ 8x_2 - 8x_2^2)^2 + (3-4x_2)^2 + 16 \sqrt{x_3 + 1}(2x_3 - 1)^2 + 30 \ln (1+x_3), \; \bm{x} \in [0,1]^3,
\label{eq:detpep}
\end{equation}
\vspace{-0.2cm}
taken from the 8-d function in \cite{DP2010}. For both functions, $N_l = 5$ levels are used for each factor, with levels set at the middle of equi-spaced intervals\footnote{Here and in later simulations, we discretized continuous test functions to provide a test bank for the discrete problem; the goal is not to solve the underlying continuous optimization problem via discretization.}. The design $\mathcal{D}_n$ used is an OA \citep{Hea2012} with randomized level permutations.

Table \ref{tbl:toill} shows the predicted minimum for $f$ using AM and PW, averaged over 100 replications. For the Friedman function, AM is a much better method than PW, yielding near-optimal settings even with a small number of evaluations $n$. This is not surprising, since the Friedman function is additive with small interactions. In the limiting case where all settings are observed (i.e., $n = \#\{\mathcal{X}\}$), both methods arrive at the global minimum of 1.81. For the DetPep10 function, PW provides improved minimization to AM for \textit{all} sample sizes $n$. This can be explained by the non-smooth nature of the DetPep10 function, which has many interaction effects. In fact, even with all settings observed (i.e., $n = \#\{\mathcal{X}\}$), AM returns a solution far away from the global minimum 10.83. This shows that both AM and PW are not effective robust optimization methods when little is known on the objective $f$.

As shown in Section \ref{sec:sim}, other existing methods (e.g., the discretized EI method) suffer from a similar lack of robustness. This is problematic in practice, since practitioners have little prior knowledge on the black-box objective function in real-world problems. To address this, the proposed ATM method (presented next) employs an adaptive trade-off between the model-based AM and the rank-based PW to achieve the desired robustness in optimization.

\section{Analysis-of-marginal-Tail-Means}
We now present the Analysis-of-marginal-Tail-Means (ATM) method, which uses \textit{conditional} tail means as marginal statistics for optimization. We first review conditional tail means, then show why and how these conditional means allow for robust optimization in ATM.

\subsection{Conditional tail means}
\textit{Conditional tail means} (or simply tail means) were first popularized in the risk management literature, and are widely used as a measure of risk for financial losses. For a scalar random variable $Z \sim F$ with finite mean, the conditional tail mean of $Z$ at percentage $\alpha \in [0,1]$ is \textit{traditionally} defined as $\mathbb{E}[Z|Z>Q_\alpha(F)]$, the expected value of $Z$ given it exceeds its 100$\alpha$\% quantile $Q_\alpha(F)$ \citep{Har2006}. When $Z$ is an (uncertain) financial loss to be incurred, the 100$\alpha$\% tail mean of $Z$ represents the expected loss in the worst-case $100(1-\alpha)\%$ event -- the event that this loss falls within the upper $100(1-\alpha)\%$ tail. Tail means are used in many aspects of risk analysis, including portfolio valuation \citep{Har2006}, options pricing \citep{Bar2016}, and as a standard for capital measurement \citep{BF2014}. 

Here, we employ a \textit{modified} version of tail means as marginal statistics for robust optimization. Let $\bm{z} = (z_1, \cdots, z_m)$ be a vector of $m$ observed values, with $z_{(1)} \leq \cdots \leq z_{(m)}$ its order statistics. For a fixed percentage $\alpha \in [0,1]$, we define the 100$\alpha$\% tail mean of $\bm{z}$ as:
\begin{equation}
\begin{cases}
\frac{1}{\lceil m\alpha \rceil} \sum_{r=1}^{\lceil m\alpha \rceil} z_{(r)}, & \alpha \in (0,1],\\
z_{(1)}, & \alpha = 0.
\end{cases}
\label{eq:cte}
\end{equation}
In other words, the 100$\alpha$\% tail mean of $\bm{z}$ is the mean of all observed values \textit{below} its 100$\alpha$\% quantile. This differs slightly from the traditional tail mean in risk analysis, which considers the mean \textit{above} a certain quantile; such a modification is needed for minimizing $f$.

From a robust statistics perspective, the conditional tail mean \eqref{eq:cte} can be viewed more generally as an L-estimator \citep{Hub2011} -- a linear combination of order statistics, which has been widely used in robust estimation. We show next how the same tail means, when used as marginal statistics, enable robust optimization as well.

% the observed function slice at level $x_l$ of factor $l$, with $m = \#\{\hat{\mathcal{F}}_l(x_l)\}$ the number of samples in this function slice, and $y_{(1)}, \cdots, y_{(m)}$ the sorted observations (from smallest to largest). (Note that each level of each factor receives an equal sample size $m$, because an orthogonal array is used for $\mathcal{D}_n$.) Suppose each factor $l$ is assigned a percentage $\alpha_l \in [0,1]$, $l = 1, \cdots, p$. The \textit{marginal tail mean} for factor $l$ at level $x_l$ is then defined as:
%\begin{equation}
%\hat{m}_{\alpha_l}(x_l) := \mathcal{M}_{{\rm tail}, \alpha_l} \{ \hat{\mathcal{F}}_l(x_l) \} = \begin{cases}
%\frac{1}{\lceil m\alpha_l \rceil} \sum_{r=1}^{\lceil m\alpha_l \rceil} y_{(r)}, & \alpha_l \in (0,1],\\
%y_{(1)}, & \alpha_l = 0,
%\end{cases}
%\label{eq:cte}
%\end{equation}
%where $\lceil \cdot \rceil$ is the ceiling function. In other words, $\hat{m}_{\alpha_l}(x_l) = \mathcal{M}_{{\rm tail}, \alpha_l}\{\hat{\mathcal{F}}_l(x_l)\}$ returns the mean of the function values in $\hat{\mathcal{F}}_l(x_l)$, conditional on such values being \textit{on} or \textit{below} the $\alpha_l$-th quantile of $\hat{\mathcal{F}}_l(x_l)$. This is slightly different from the earlier financial definition of tail means; here, the interest lies in the \textit{lower} tail of the distribution.

\subsection{Optimization with marginal tail means}

\begin{figure}[t]
\begin{minipage}{0.45\textwidth}
\centering
\includegraphics[width=0.9\textwidth]{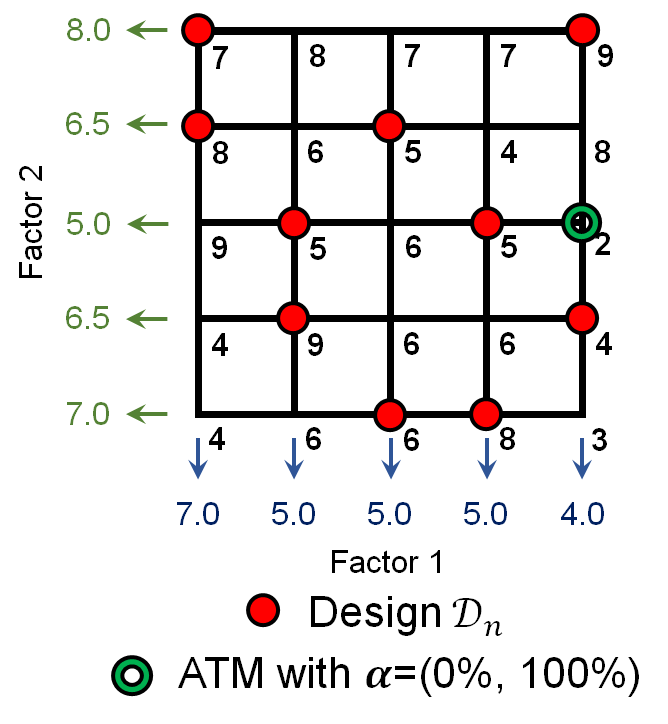}
\captionof{figure}{A visualization of ATM with $\boldsymbol{\alpha} = (0\%,100\%)$ for a 2-d toy example. Blue and green arrows show the estimated marginal tail means for factors 1 and 2.}
\label{fig:atm}
\end{minipage}
\hspace{0.2cm}
\begin{minipage}{0.54\textwidth}
\centering
\includegraphics[width=1.02\textwidth]{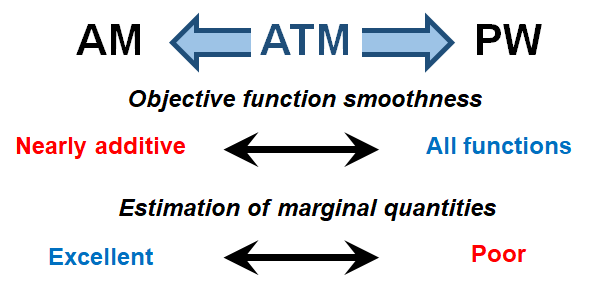}
\captionof{figure}{Two implicit trade-offs enabling robust optimization in ATM.}
\label{fig:tradeoff}
\end{minipage}
\end{figure}

ATM uses tail means for optimization in the following way. First, assign to each factor $l$ a percentage $\alpha_l \in [0,1]$, and let $m_{\alpha_l}(x_l)$ be the true marginal 100$\alpha_l$\% tail mean of factor $l$ at level $x_l \in [N_l]$, with $\hat{m}_{\alpha_l}(x_l)$ its estimate from data. The ATM predictor of $\bm{x}^*$ becomes:
\begin{equation}
\hat{\bm{x}}_{\boldsymbol{\alpha}} = (\hat{x}_1, \cdots, \hat{x}_p), \quad \hat{x}_l = \underset{x_l \in [N_l]}{\text{argmin}} \; \hat{m}_{\alpha_l}(x_l), \quad l = 1, \cdots, p.
\label{eq:atm}
\end{equation}
In other words, ATM selects the level $\hat{x}_l$ with smallest estimated marginal \textit{tail} mean for each factor $l$, then uses the level combination $\hat{\bm{x}}_{\boldsymbol{\alpha}} = (\hat{x}_1, \cdots, \hat{x}_p)$ to predict $\bm{x}^*$.

Figure \ref{fig:atm} visualizes ATM using the earlier 2-d toy example. Suppose the percentages $\boldsymbol{\alpha} = (\alpha_1,\alpha_2) = (0\%, 100\%)$ are chosen for the two factors. The blue and green arrows show the estimated marginal tail means for the two factors: the $0$\% tail mean (i.e., the marginal minimum) for factor 1, and the $100$\% tail mean (i.e., the marginal mean) for factor 2. For factor 1, the $\hat{x}_1=5$-th level has the smallest tail mean, whereas for factor 2, the $\hat{x}_2=3$ has the smallest tail mean. From \eqref{eq:atm}, the ATM prediction of the optimal setting $\bm{x}^*$ is $\hat{\bm{x}}_{\boldsymbol{\alpha}} = (5,3)$, with objective value $f(\hat{\bm{x}}_{\boldsymbol{\alpha}}) = 2$ lower than both the AM and PW predictions.

One appealing property of ATM is that it bridges between the purely model-based AM and the purely rank-based PW methods. To see this, set the percentages $\alpha_l = 100\%$ for all factors $l$. The resulting ATM predictor $\hat{\bm{x}}_{\boldsymbol{\alpha}}$ then selects levels based on the estimated (unconditional) marginal means, which is precisely AM. Similarly, setting $\alpha_l = 0\%$ for all factors, the ATM predictor $\hat{\bm{x}}_{\boldsymbol{\alpha}}$ selects levels based on the estimated marginal minima. But the level with the smallest estimated marginal minimum must contain the smallest observed response, so ATM reduces to PW. In this sense, ATM with percentages $\alpha_l \in (0,1)$ integrates both rank-based and model-based characteristics for optimization, as advocated in \cite{Wea1987} (see Section \ref{sec:intro}). This integration enables robust optimization via two trade-offs on objective function smoothness and marginal estimation, which we describe below.

%\begin{figure}[t]
%
%\end{figure}

%It is still unclear what specific properties are being traded off between AM and PW. To this end, we present below three fundamental trade-offs concerning response surface smoothness, the estimation accuracy of marginal quantities, and the dichotomy of model-based vs. rank-based optimization (see Figure \ref{fig:tradeoff} for a visualization), and show how the proposed tail means in ATM nicely parameterize such trade-offs.

\subsection{Two trade-offs for robustness}
\label{sec:tradeoff}
There are two implicit trade-offs in ATM which allow for effective robust optimization. The first trade-off is on objective function smoothness, and the second trade-off is on estimation accuracy of marginal statistics. Figure \ref{fig:tradeoff} shows a visualization of these trade-offs.

\subsubsection{Trade-off 1: Objective function smoothness}
\label{sec:smooth}

Recall from Section \ref{sec:ampwex} that AM performs well for the nearly additive Friedman function, but is inferior to PW for the more non-smooth DetPep10 function, which suggests a difference in the assumed objective smoothness between AM and PW. The following proposition formalizes this by showing, in the limiting sense, AM returns the optimal setting $\bm{x}^*$ for a particular type of ``nearly additive'' objective function:
\begin{proposition}[Convergence of AM]
Suppose all settings in $\mathcal{X}$ are observed with no noise. Let $\hat{\bm{x}}_{\rm AM} = (\hat{x}_1, \cdots, \hat{x}_p)$ be the AM predictor in \eqref{eq:am}. If, for all factors $l = 1, \cdots, p$ and $\bm{x}_{-l} \in \mathcal{X}_{-l}$, $f$ satisfies:
\begin{equation}
f(\hat{x}_l; \bm{x}_{-l}) \leq f(x_l;\bm{x}_{-l}), \quad x_l \in [N_l],
\label{eq:fnform}
\end{equation}
then $\hat{\bm{x}}_{\rm AM} = \bm{x}^*$. Here, $\mathcal{X}_{-l}$ is the full factorial space $\mathcal{X}$ without factor $l$.
\label{thm:amc}
\end{proposition}
\noindent In words, if $f$ satisfies condition \eqref{eq:fnform}, Proposition \ref{thm:amc} guarantees that AM converges to $\bm{x}^*$ when all settings are observed without noise. The proof of Proposition \ref{thm:amc} is in the Appendix.

\begin{figure}[t]
\begin{minipage}{0.45\textwidth}
\centering
\includegraphics[width=0.77\textwidth]{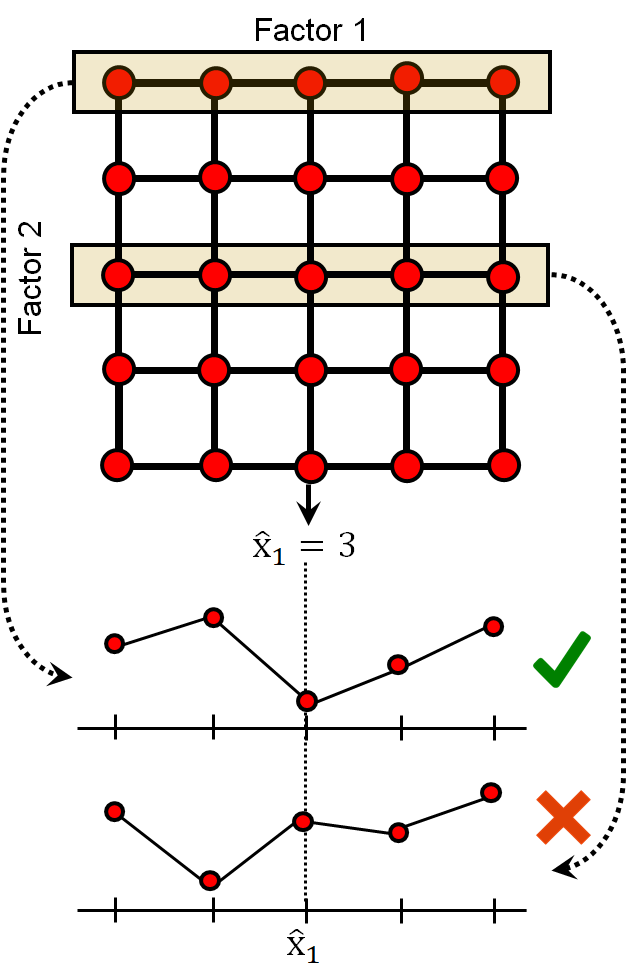}
\captionof{figure}{A visualization of when the MC condition in \eqref{eq:fnform} is satisfied or violated.}
\label{fig:margcond}
\end{minipage}
\hspace{0.2cm}
\begin{minipage}{0.54\textwidth}
\centering
\includegraphics[width=0.82\textwidth]{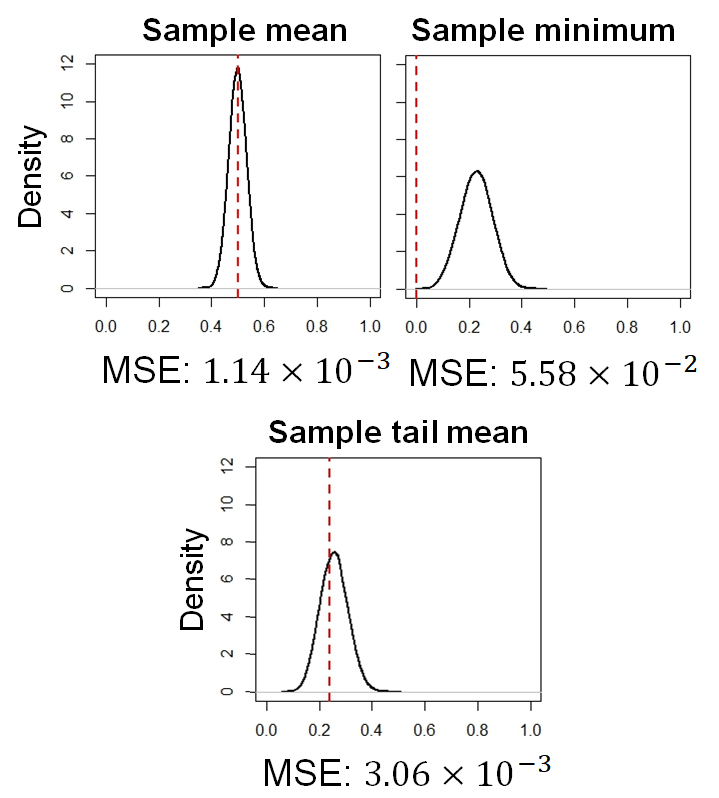}
\captionof{figure}{Sampling distributions for the sample mean, sample minimum and sample $20$\% tail mean. Dotted red lines indicate true parameters. The MSE of each estimator is shown on the bottom.}
\label{fig:dens}
\end{minipage}
\end{figure}

Condition \eqref{eq:fnform}, which we call the \textit{marginal-conditional} (MC) requirement, shows which functions can be optimized well using AM. This requirement can be explained as follows: for each factor $l$, every function slice of $f$ \textit{conditional} on other factors $\bm{x}_{-l}$ must be minimized at $\hat{x}_l$, the level with the smallest \textit{marginal} mean. Figure \ref{fig:margcond} visualizes this using a 2-d example. Suppose $\hat{x}_1 = 3$, i.e., factor 1 has the smallest marginal mean at level 3. The two plots on the bottom show two function slices of $f$, given fixed levels of the other factor (factor 2). For the first slice, its minimum is found at level $3$, which matches the level with smallest marginal mean $\hat{x}_1 = 3$. For the second slice, however, its minimum is found at level 2, which does not match $\hat{x}_1 = 3$. Because of this, $f$ does not satisfy the MC condition, and AM is not expected to perform well here. The MC condition can be seen as a generalization of the monotonicity condition in \cite{Wea1987}, which requires $f$ to be of the form:
\begin{equation}
f(\bm{x}) = \psi\{f_1(x_1), \cdots, f_p(x_p)\}, \quad f_l(x_l) = \frac{1}{\# \{ \mathcal{X}_{-l}\} } \sum_{\bm{x}_{-l} \in \mathcal{X}_{-l}} f(x_l; \bm{x}_{-l}),
\label{eq:mono}
\end{equation}
where $\psi$ is a non-decreasing function in each argument, and $f_l$ is the $l$-th main effect of $f$.

When $f$ is additive (i.e., of the form \eqref{eq:add}), the MC condition must hold, since all conditional slices equal the marginal mean values plus some constant. Even when $f$ is non-additive with small interactions (e.g., the Friedman function), this condition holds as long as these interactions do not change the minimum level of each conditional slice, and AM should optimize such functions well from Proposition \ref{thm:amc}. However, the MC condition is easily violated if $f$ has moderate interactions (e.g., DetPep10). For such functions, AM can return poor solutions even when \textit{all} settings are observed, as seen in Section \ref{sec:ampwex}.

PW, on the other hand, requires no such structure on $f$. Suppose again that all settings are observed with no noise. In this case, the PW predictor $\hat{\bm{x}}_{\rm PW}$ always arrives at the optimal setting $\bm{x}^*$, since it picks the winner from observed settings. Compared to AM, PW enjoys a stronger guarantee in terms of optimization convergence, since it always arrives at $\bm{x}^*$ (albeit not very efficiently) without any assumptions on $f$.

With percentages $\alpha_l \in (0,1)$, ATM trades-off between the restricted function space of AM and the unrestricted space of PW. To see why, suppose $\alpha_l = 50\%$ for all factors $l=1, \cdots, p$. The ATM predictor $\hat{\bm{x}}_{\boldsymbol{\alpha}}$ in \eqref{eq:atm} can be seen as a two-step procedure. First, the largest $\alpha_l=50\%$ of the data are removed from each level of a factor (ranking). Second, using the remaining data, the levels with the smallest marginal means are selected (modeling). These two steps implicity assume the objective function $f$ to be \textit{locally} additive around its minimum -- it first identifies a local region by ranking and removing the largest $(1-\alpha_l)\%$ observations within each level, then predicts using a main-effects model on remaining data. With $\alpha_l = 100\%$, this reduces to the (global) additive model in AM, and with $\alpha_l=0\%$, this generalizes to the unrestricted function space for PW.

\subsubsection{Trade-off 2: Estimation accuracy of marginal quantities}
\label{sec:acc}

The second trade-off in ATM concerns the estimation accuracy of marginal tail means with limited data. Consider the two extremes of AM ($\alpha_l = 100\%$) and PW ($\alpha_l = 0\%$), which use means and minima as marginal statistics for optimization. On one hand, the marginal means for AM can be well estimated with limited runs, given the use of a balanced design which allocates multiple observations for each factor level. On the other hand, the marginal minima in PW are difficult to estimate even with many observations, because the sample minimum is always positively biased. The marginal tail means in ATM bridge between the excellent estimability of marginal statistics in AM and the poor estimability for PW.

To better understand this trade-off, consider the following toy example. Suppose data $\bm{z} = (z_1, \cdots, z_m)$ is sampled i.i.d. from a $\text{Beta}(5,5)$ distribution, and the goal is to estimate its mean, minimum, and $20\%$ tail mean with $m=20$ samples. Figure \ref{fig:dens} shows the sampling distributions of the sample mean, minimum, and 20\% tail mean, along with their mean-squared errors (MSEs) in estimating the true distributional quantities (dotted in red). The sample mean, as expected, yields unbiased estimates and enjoys low estimation errors. The sample minimum, on the other hand, has large positive bias and high variance, resulting in poor estimation in terms of MSE. The sample tail mean strikes a compromise between these two extremes, providing good estimation with a slight positive bias.

%If PW offers the best optimization guarantees, why then does AM perform better than PW for the Friedman function? The second trade-off offers an answer to this question: the marginal means in AM can be estimated more accurately than the marginal minima in PW.

%\begin{figure}[t]
%\end{figure}

%To see this, suppose the marginal samples in $\hat{\mathcal{F}}_l(x_l)$ are sampled uniformly-at-random from the function slice $\mathcal{F}_l(x_l)$, with the function values in $\mathcal{F}_l(x_l)$ randomly sampled from a $Beta(5,5)$ distribution (this ensures values are concentrated near the median; see left plot in Figure \ref{fig:dens}). The right plots in Figure \ref{fig:dens} shows the sampling distributions of the estimated marginal quantity $\hat{m}_l(x_l)$ for the mean, minimum and tail mean (with $\alpha = 0.2$), using $m=20$ samples. The dotted red lines indicate the true marginal quantity $m_l(x_l)$ (i.e., mean, minimum or tail mean), with the corresponding mean-squared-errors (MSE) of the sampling distributions shown below. For the marginal mean, we see that the sample estimator $\hat{m}_l(x_l)$ is unbiased, and enjoys low estimation error in terms of MSE. For the marginal minimum, however, the sample estimator $\hat{m}_l(x_l)$ has a large positive bias and high variance, thereby yielding poor estimation in terms of MSE. The proposed tail mean estimator strikes a compromise between the excellent estimation of the marginal mean with the poor estimation of the marginal minimum.

Putting these two trade-offs together, we get a complete view on why ATM is an effective robust optimization method. By integrating both rank- and model-based optimization methods, ATM can adaptively set its percentages $\boldsymbol{\alpha}$ to identify an additive region near the minimum, then exploit this additivity for prediction via marginal tail means. When $f$ is nearly additive, ATM with $\alpha_l \approx 100\%$ fully exploits this global structure to effectively predict the optimal setting with limited runs. When $f$ is non-additive with strong interactions, ATM with $\alpha_l \approx 0\%$ relies on the ranking of observed settings to guide optimization. Finally, when $f$ is non-additive with moderate interactions, ATM with a moderate choice of $\alpha_l$ exploits local additive structure to predict $\bm{x}^*$. This flexible framework, when tuned using data, enables ATM to effectively optimize a broad range of objective functions.

\subsection{Back to motivating examples}
\label{sec:mot}

\begin{figure}[t]
\centering
\includegraphics[width=0.42\textwidth]{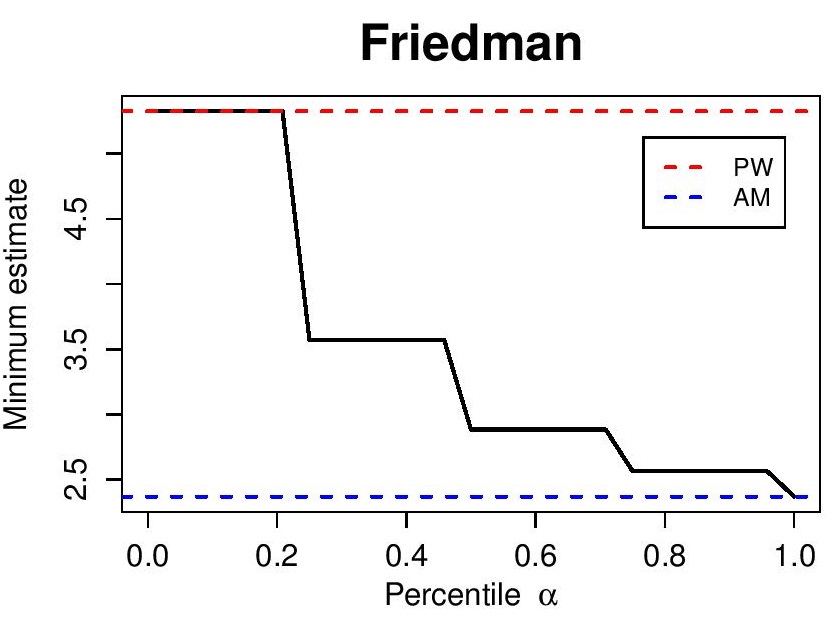}
%\hfill
\includegraphics[width=0.42\textwidth]{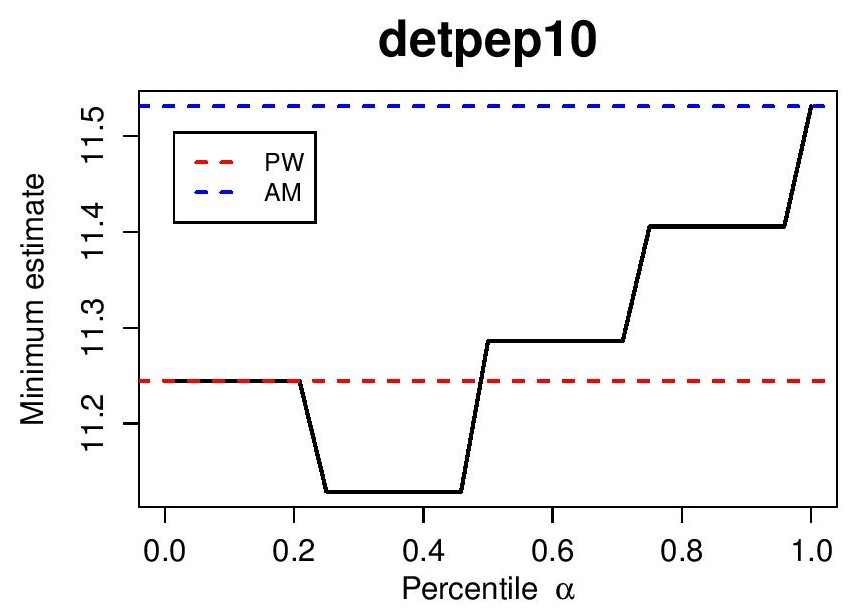}
\caption{Mean predicted minimum $f(\hat{\bm{x}}_{\boldsymbol{\alpha}})$ using ATM, averaged over 100 randomly-permutated OAs with $n=25$ function evaluations. The blue and red dotted lines indicate AM (ATM with $\alpha = 100\%$) and PW (ATM with $\alpha = 0\%$), respectively.}
\label{fig:motcte}
\end{figure}

To illustrate ATM, let us return to the earlier examples. Consider the simple case of $\alpha_l = \alpha$, i.e., the same percentage $\alpha$ for all $p$ factors. Figure \ref{fig:motcte} shows the predicted minimum value $f(\hat{\bm{x}}_{\boldsymbol{\alpha}})$ using the ATM predictor $\hat{\bm{x}}_{\boldsymbol{\alpha}}$ for different choices of $\alpha \in (0,1)$. Note that the two endpoints of $\alpha = 100\%$ and $\alpha = 0\%$ (dotted in blue and red) correspond to AM and PW, respectively. For Friedman, ATM gives better predictions as $\alpha$ increases, with $\alpha=100\%$ (i.e., AM) providing the lowest objective value. This makes sense in terms of the two trade-offs: the Friedman function is nearly additive (i.e., it satisfies the MC condition, trade-off 1), so a larger $\alpha$ enables better estimation of marginal quantities (trade-off 2) and hence better minimization. The results for DetPep10 are more interesting: ATM with $\alpha = 0\%$ (i.e., PW) performs better than ATM with $\alpha = 100\%$ (i.e., AM), but ATM with $\alpha \approx 30\%$ improves upon both. This can again be explained using the two trade-offs: the DetPep10 function is non-smooth with moderate interactions (i.e., it violates the MC condition, trade-off 1), so a choice of $\alpha \in (0,1)$ allows ATM to exploit local additive structure via marginal tail means (trade-off 2) for effective optimization. We introduce next a tuning procedure for $\boldsymbol{\alpha}$, which uses observed data to estimate a good compromise between these two trade-offs.

%This illustrates a key advantage of ATM: when MCR is violated for the response surface $f$, ATM offers a way to trade-off between function smoothness and estimation accuracy via the choice of percentages $\boldsymbol{\alpha}$, with the optimization performance from an optimal choice of $\boldsymbol{\alpha}$ potentially better than both AM and PW.

\section{Implementing ATM}
\label{sec:impl}
Next, we discuss implemention steps for ATM, beginning with a tuning procedure for percentages $\boldsymbol{\alpha}$, then a level elimination strategy for batch-sequential optimization, and finally some comments on experimental design.
%A practical problem still remains: how does one tune for an optimal choice of percentages $\boldsymbol{\alpha}$, using \textit{only} the data observed at design $\mathcal{D}_n$? We present in this section an effective tuning method which makes use of a fitted surrogate model on $f$, then introduce a sequential implementation of ATM which employs bootstrapping and level elimination to guide the optimization procedure.

\subsection{\texttt{tune.alpha} -- Tuning ATM percentages}
\label{sec:alpha}
We present here a method, called \texttt{tune.alpha}, for tuning ATM percentages $\boldsymbol{\alpha}$ from data. Recall the two trade-offs earlier: $\boldsymbol{\alpha}$ should be set as large as possible to exploit marginal structure, but not so large as to violate the assumed smoothness on $f$. \texttt{tune.alpha} tunes a good balance between these trade-offs using two steps: it first (i) uses observed data to fit a model $\hat{f}$, then (ii) employs $\hat{f}$ to estimate a good choice of percentages $\hat{\boldsymbol{\alpha}}$ for optimization. 

Consider first step (i), which fits a simple model $\hat{f}$ with main effects and two-factor interactions. To fit this model, a state-of-the-art method called \texttt{hierNet} \citep{Bea2013} is used, which employs convex optimization to select and estimate important effects in $f$. This fitted model $\hat{f}$ has the desirable property of (weak) effect heredity \citep{WH2009}, meaning all selected interactions have at least one active main effect component. Consider next step (b), which uses the fitted model $\hat{f}$ to tune ATM percentages $\boldsymbol{\alpha}$. The strategy is to find percentages $\hat{\boldsymbol{\alpha}}$ which, when used within ATM, yield the smallest predicted value on the \textit{fitted} model $\hat{f}$. However, the same dataset $(\mathcal{D}_n,\bm{y})$ for fitting $\hat{f}$ in step (i) should not be used to tune $\boldsymbol{\alpha}$ in step (ii), since this introduces unwanted bias in tuning. Instead, we perform this tuning step on a synthetic dataset $(\mathcal{D}^*,\bm{y}^*)$, where $\mathcal{D}^*$ is a new OA randomly sampled on $\mathcal{X}$, and $\bm{y}^* = \hat{f}(\mathcal{D}^*)$ are its observations generated from the \textit{fitted} model $\hat{f}$. Algorithm \ref{alg:tunealpha} outlines the detailed steps for \texttt{tune.alpha}.

To understand why these two steps enable effective tuning of $\boldsymbol{\alpha}$, suppose the true function $f$ is nearly additive. For such functions, true interaction effects are weak relative to true main effects, so the \textit{fitted} interactions of $\hat{f}$ from \texttt{hierNet} should be weak relative to \textit{fitted} main effects as well. In this case, the tuned percentages $\hat{\boldsymbol{\alpha}}$ from \texttt{tune.alpha} will all be close to 100\%, because the nearly additive fitted model $\hat{f}$ is best optimized with AM. On the other hand, suppose $f$ has large interactions relative to main effects. In this case, \texttt{hierNet} should be able to detect and reflect the large interactions-to-main-effects ratio in the \textit{fitted} model $\hat{f}$. The tuned percentages $\hat{\boldsymbol{\alpha}}$ will then shift closer to 0\%, to identify a local additive region near the minimum for prediction. For high-dimensional problems, the exact optimization for $\hat{\boldsymbol{\alpha}}$ can be computationally demanding; we find that a simple Monte Carlo approximation (with $\boldsymbol{\alpha} = \bm{0}$ and $\boldsymbol{\alpha} = \bm{1}$ included as candidate points) works quite well.
%The tuned ATM predictor $\hat{\bm{x}}_{\hat{\boldsymbol{\alpha}}}$ (with tuned percentages $\hat{\boldsymbol{\alpha}}$) is then used to estimate the minimum setting for the \textit{true} surface $f$.

%Implementation-wise, \texttt{tune.alpha} performs the following three steps. First, using the function evaluations from design $\mathcal{D}_n$, we fit a surrogate surface $\hat{f}$ using \texttt{hierNet}, which identifies and estimates important main effects and interactions in $f$ using convex programming.  Next, using $\hat{f}$, we search for the optimal percentages $\hat{\boldsymbol{\alpha}}$ from a grid of candidate percentages $\mathcal{G}$, so that $\hat{f}(\hat{\bm{x}}_{\hat{\boldsymbol{\alpha}}}) \leq \hat{f}(\hat{\bm{x}}_{\boldsymbol{\alpha}})$ for all candidate percentages $\boldsymbol{\alpha} \in \mathcal{G}$. Finally, the tuned percentages $\hat{\boldsymbol{\alpha}}$ are then incorporated into the ATM predictor $\hat{\bm{x}}_{\hat{\boldsymbol{\alpha}}}$ for estimating $\bm{x}^*$.

We clarify a key distinction between ATM (tuned using \texttt{tune.alpha}) and a direct minimization of the fitted model $\hat{f}$. The first uses the relative strength of interactions to main effects in $\hat{f}$ to \textit{tune} a good choice of ATM percentages $\hat{\boldsymbol{\alpha}}$, whereas the second relies fully on the fitted model $\hat{f}$ for minimization. As noted in \cite{Wea1990}, the latter strategy may yield poor results when the true surface $f$ is non-smooth, because $\hat{f}$ can fit $f$ poorly around the desired minimum. To contrast, \texttt{tune.alpha} employs the relative strength of fitted interactions to main effects (which can be more reliably estimated) to tune a good minimization strategy within ATM. This adaptive tuning of $\boldsymbol{\alpha}$ from data allows ATM to effectively optimize a wide range of black-box problems with limited data, as shown later.

\begin{algorithm}[t]
\caption{\texttt{tune.alpha}$(\mathcal{D}_n, \bm{y})$: Tuning ATM percentages}
\label{alg:tunealpha}
\begin{algorithmic}
\stb Using data $(\mathcal{D}_n,\bm{y})$, fit a second-order interactions model $\hat{f}$ with \texttt{hierNet}.
\stb Randomly generate a new OA $\mathcal{D}^*$ on $\mathcal{X}$ with $\# \{\mathcal{D}^*\} \leq n$, and set $\bm{y}^* \leftarrow \hat{f}(\mathcal{D}^*)$.
\stb Set $\hat{\boldsymbol{\alpha}} \leftarrow \argmin_{\boldsymbol{\alpha} \in [0,1]^p} \hat{f}(\hat{\bm{x}}^*_{\boldsymbol{\alpha}})$, where $\hat{\bm{x}}^*_{\boldsymbol{\alpha}}$ is the ATM predictor using data $(\mathcal{D}^*,\bm{y}^*)$.
\stb Return $\hat{\boldsymbol{\alpha}}$.
\end{algorithmic}
\end{algorithm}

\subsection{\texttt{sel.atm} -- Sequential-Elimination-of-Levels using ATM}
\label{sec:sel}
Next, we introduce an iterative implementation of ATM, called \texttt{sel.atm}, for batch-sequential optimization. This extension of ATM is important for practical problems, where experiments are often run in sequential batches due to process limitations or computing architecture.

\texttt{sel.atm} adopts the Sequential-Elimination-of-Levels (SEL) framework in \cite{Wea1990}. The key idea behind SEL is to employ batch-sequential sampling to iteratively remove worst performing levels of each factor, which enables experimenters to quickly target regions of interest with few samples. Applied to AM and PW, SEL iteratively removes from each factor the level with the largest marginal mean and minimum, respectively (these methods are called \texttt{sel.mean} and \texttt{sel.min}). Here, instead of using marginal means or minima, \texttt{sel.atm} sequentially removes levels with the largest marginal \textit{tail} means tuned from data. This can be interpreted as iteratively removing the worst levels from an additive model around the minimum. When these tail means are tuned from data via \texttt{tune.alpha}, \texttt{sel.atm} can provide a \textit{robust} elimination of factor levels.

\begin{algorithm}[t]
\caption{\texttt{sel.atm}$(T_{\rm elim})$: Sequential-Elimination-of-Levels using ATM}
\label{alg:selatm}
\begin{algorithmic}
\stb Initialize design $\mathcal{D}_n \leftarrow \varnothing$ and observations $\bm{y} \leftarrow \varnothing$.
\For{$t = 1, \cdots, T_{\rm elim}$} \Comment{$T_{\rm elim}$ -- Number of level eliminations}
\stb Add to $\mathcal{D}_n$ the smallest OA with levels $\prod_{l=1}^p (N_l - t + 1)$, and add to $\bm{y}$ the observations collected on this OA.
\stb Tune ATM percentages: $\hat{\boldsymbol{\alpha}} \leftarrow \text{\texttt{tune.alpha}}(\mathcal{D}_n,\bm{y})$.
%\For{$l=1, \cdots, p$ and $x_l \in [N_l]$} \Comment{For each factor $l$ and each level $x_l$...}
%\stb Obtain the de-biased estimates $\hat{m}_{\hat{\alpha}_l}(x_l) \leftarrow \text{\texttt{super.boot}}(\hat{\mathcal{F}}_l(x_l), \mathcal{M}_{{\rm tail}, \hat{\alpha}_l}, R_{\rm boot})$.
%\EndFor
\stb Predict minimum with $\hat{\bm{x}}_{\hat{\boldsymbol{\alpha}}}$, the ATM predictor using data $(\mathcal{D}_n,\bm{y})$.
\For{$l=1, \cdots, p$} \Comment{For each factor $l$...}
\stb Remove level $x_l$ with largest marginal tail mean estimate $\hat{m}_{\hat{\alpha}_l}(x_l)$.
\EndFor
\EndFor
%\stb Return predicted optimal setting $\hat{\bm{x}}_{\hat{\boldsymbol{\alpha}}}$.
\end{algorithmic}
\end{algorithm}

\texttt{sel.atm} consists of the following steps. First, data $\bm{y}$ are collected from experiments using an OA design, and the tail mean percentages $\boldsymbol{\alpha}$ are tuned via \texttt{tune.alpha}. Next, the tuned percentages $\hat{\boldsymbol{\alpha}}$ are plugged into the ATM predictor $\hat{\bm{x}}_{\hat{\boldsymbol{\alpha}}}$, to be used as the estimate of the optimal setting $\bm{x}^*$. Finally, for each factor $l$, the level $x_l$ with the largest marginal tail mean $\hat{m}_{\hat{\alpha}_l}(x_l)$ is eliminated. These steps are repeated until a unique setting remains, or for a fixed number of eliminations $T_{\rm elim}$. Algorithm \ref{alg:selatm} summarizes the above steps for \texttt{sel.atm}.

%The original motivation behind the SEL method is that one can make use of marginal statistics, such as the mean or the minimum, to sequentially eliminate unwanted factor levels. However, from Section \ref{sec:tradeoff}, we know that SEL using the marginal mean can eliminate the wrong levels when $f$ violates MCR; likewise, SEL with the marginal minimum can erroreously remove levels due to poor marginal estimation. Viewed this way, \texttt{sel.atm} offers an adaptive scheme which uses the observed data to tune an optimal trade-off between function smoothness and estimation accuracy, thereby providing a more effective and robust level elimination method for optimizing both smooth and rugged response surfaces.

\subsection{Experimental design}
\label{sec:des}
For \texttt{sel.atm} (or ATM), which relies on marginal information for optimization, the choice of experimental design (on discrete space $\mathcal{X}$) can impact its performance. A good design should have two properties. First, it should be \textit{balanced} for estimating main effects, in that all factor levels are observed an equal number of times -- this prevents ATM from being skewed towards any level of a factor. Second, a design should enable good estimation of interaction effects, which in turn allow for effective tuning of percentages $\boldsymbol{\alpha}$.

Given these two properties and the need for small run sizes, we find that orthogonal arrays \citep{Hea2012} provide good designs for \texttt{sel.atm}: its first-order balance ensures good main effect estimation, whereas its second-order balance ensures good estimation of two-factor interactions. OAs are also flexible designs which accommodate mixed-level problems (i.e., problems with different number of levels $N_l$) with a small run size -- these are called \textit{mixed-level} OAs \citep{WH2009}. This flexibility is important in practice: some factors may be binary (e.g., on or off) while others can have more levels (e.g., parts A, B or C). We will use OAs as the design of choice for \texttt{sel.atm} in later applications.

It is important to clarify here why OAs are preferred for \texttt{sel.atm} compared to more standard designs for black-box continuous optimization, such as Latin hypercube designs \citep{Mea1979} or space-filling designs \citep{Jea1990}. The key reason is that we are interested in optimizing over a \textit{discrete} factorial space $\mathcal{X}$. In the special case where \textit{all} factors are discretized from a controllable, continuous scale, the latter designs are useful for optimization on the underlying continuous space. However, when some factors are nominal or discretized due to experimental constraints, the latter designs cannot be used. OAs provide a more flexible design scheme for the broader discrete problem at hand, which accommodates all types of discrete factors.

Another point of consideration is design run size. Given the expensive nature of experiments, the number of runs should be kept as small as possible, while ensuring a good solution from \texttt{sel.atm} (or ATM). One way to do this is to use the \textit{smallest} OA design within each stage of \texttt{sel.atm} -- this ensures a small run size while retaining the desired OA structure. Such an approach was employed in \cite{Wea1990} and other following works. Given a \textit{fixed} run budget which exceeds this minimum run size, one can also run larger OAs in one or more stages of \texttt{sel.atm}, particularly in stages where noticeable non-additivity is detected. This is explored further in the next section. Remaining runs can also be used to explore settings near the predicted minimum.

\section{Simulation studies}
\label{sec:sim}

\subsection{Set-up}
We now compare in simulations the performance of \texttt{sel.atm} with existing discrete, black-box optimization methods. The focus is on investigating \textit{robustness}, i.e., how well a method performs over a wide range of problems. To do this, we use three test functions in the literature, each with different properties. The first is the (modified) exponential function (DetPep10e) from \cite{DP2010}:
\begin{equation}
f(\bm{x}) = 100 \sum_{k=1}^{p/3} \left( e^{ -2/{x_{3k-2}^{1.75}} } + e^{-2/{x_{3k-1}^{1.5}} } + e^{ -2/{x_{3k}^{1.25}} } + 0.01 x_{3k-2} x_{3k-1} x_{3k} \right), \quad \bm{x} \in [0,1]^p,
\label{eq:dpe}
\end{equation}
a nearly additive function with a small three-way interaction. The second is the six-hump camel function (Camel6) from \cite{Aea2005}:
\begin{equation}
\small
f(\bm{x}) = \sum_{k=1}^{p/2} \left( 4 - 2.1x_{2k-1} + \frac{x_{2k-1}^4}{3} \right) x_{2k-1}^2 + x_{2k-1}x_{2k} + (-4 + 4x_{2k}^2)x_{2k}^2, \quad \bm{x} \in ([-2, 2] \times [-1, 1])^{p/2},
\label{eq:sh}
\end{equation}
which is non-additive with moderate interactions. The last is the (modified) Shubert function \citep{JY2013}:
\begin{equation}
f(\bm{x}) = \frac{1}{10^p}\left\{\prod_{k=1}^p \left( \sum_{i=1}^5 i \cos ( (i+1) x_k + i )\right) + 0.01 \prod_{k=1}^p x_k \right\}, \quad \bm{x} \in [-10,10]^p,
\label{eq:shu}
\end{equation}
a non-additive function with strong interactions. DetPep10e is tested with $p=9$ and $18$ factors, Camel6 with $p=8$ and $24$ factors, and Shubert with $p=10$ and $24$ factors, with each factor having $N_l = 4$ levels (set at the middle of equi-spaced intervals).

Here, \texttt{sel.atm} is run with $T_{\rm elim} = 2$ level eliminations, using the smallest OA design in each stage (unless otherwise stated). For example, for the DetPep10e ($p=9$) function with $N_l = 4$ levels, the initial stage is run with a 32-run OA (the smallest run size for OA$(\cdot,4^9)$), the first elimination stage is run with a 27-run OA (the smallest run size for OA$(\cdot, 3^9)$, with one level removed), and the second stage is run with a 12-run OA (the smallest run size for OA$(\cdot, 2^9)$, with two levels removed). This corresponds to a total run size of $n=32$, $n=32+27 = 59$ and $32+27+12 = 71$ for the three stages. Run sizes for other functions are set similarly, with OAs computed using the \textsf{R} package \texttt{DoE.base} \citep{Gro2017}.

\texttt{sel.atm} is then compared with four existing methods\footnote{SELC \citep{Mea2006} and $\mathcal{G}$-SELC \citep{Mea2009} are not included here, since these methods (as implemented in \citealp{Jea2008}) require a larger run size $n$ in most simulation cases.}. The first two, \texttt{sel.mean} and \texttt{sel.min}, are the SEL schemes in \cite{Wea1990}, which employ AM and PW (see Section \ref{sec:sel}). The last two are variants of the EI method \citep{Jea1998}, modified for the discrete problem at hand. The first variant of EI (\texttt{ei.ord}) considers \textit{ordinal} discrete factors -- it uses a Gaussian process (GP) model on $\mathcal{X}$ with first-order autocovariance function:
\begin{equation}
c_{\rm ord}(\bm{x}_1, \bm{x}_2) = \sigma^2 \exp\left\{-\sum_{l=1}^p \theta_l (x_{1,l}-x_{2,l})^2\right\}, \quad \bm{x}_1, \bm{x}_2 \in \mathcal{X},
\label{eq:ordcorr}
\end{equation}
to guide optimization. The second variant of EI (\texttt{ei.nom}) considers \textit{nominal} discrete factors -- it uses a GP on $\mathcal{X}$ with exchangeable covariance function \citep{Ste2012}:
\begin{equation}
c_{\rm nom}(\bm{x}_1, \bm{x}_2) = \sigma^2 \exp\left\{ -\sum_{l=1}^p \theta_l \mathbf{1}\{x_{1,l} \neq x_{2,l}\} \right\}, \quad \bm{x}_1, \bm{x}_2 \in \mathcal{X}.
\label{eq:excorr}
\end{equation}
The scale parameters $(\theta_l)_{l=1}^p$ and variance parameter $\sigma^2$ are adaptively estimated from data via maximum likelihood. Given the prevalence of batch-sequential experiments in real-world applications (Section \ref{sec:sel}) and the discrete problem at hand, both EI methods are run at the same batch sample sizes as the first three methods, with OAs as initial designs.

%In other words, the first assumes a first-order autoregressive structure over each level of an ordinal factor, whereas the second assumes equal correlations between different levels of a nominal factor.

\begin{table}
\centering
\footnotesize
\begin{tabular}{c c c c | c c c }
\toprule
\multicolumn{1}{c}{} & \multicolumn{3}{c|}{\textbf{DetPep10e}, $p=9$} & \multicolumn{3}{c}{\textbf{DetPep10e}, $p=18$}\\
\multicolumn{1}{c}{} & $T_{\rm elim} = 0$ & $T_{\rm elim} = 1$ & $T_{\rm elim} = 2$ & $T_{\rm elim} = 0$ & $T_{\rm elim} = 1$ & $T_{\rm elim} = 2$\\
\multicolumn{1}{c}{Method} & $(n=32)$ & $(n=59)$ & $(n=71)$ & $(n=64)$ & $(n=118)$ & $(n=138)$\\
\toprule
\texttt{sel.mean} & $\crdbm{0.01}$ & \cblb{0.15} & $\crdbm{0.01}$ & $\crdbm{0.02}$ & $\crdbm{0.01}$ & $\crdbm{0.01}$ \\
\texttt{sel.min} & 3.85 & 1.21 & 0.16 & 15.5 & 6.14 & 3.14 \\
\texttt{ei.ord} & 4.05 & $\crdbm{0.06}$ & $0.04$ & 15.9 & 3.41 & 2.69\\
\texttt{ei.nom} & 3.95 & 0.26 & $0.04$ & 15.6 & 3.60 & 2.74\\
\texttt{sel.atm} & \cblb{0.37} & 0.24 & $\cblbm{0.02}$ & $\crdbm{0.02}$ & $\cblbm{0.08}$ & $\crdbm{0.01}$\\
\toprule
\multicolumn{1}{c}{} & \multicolumn{3}{c|}{\textbf{Camel6}, $p=8$} & \multicolumn{3}{c}{\textbf{Camel6}, $p=24$}\\
\multicolumn{1}{c}{} & $T_{\rm elim} = 0$ & $T_{\rm elim} = 1$ & $T_{\rm elim} = 2$ & $T_{\rm elim} = 0$ & $T_{\rm elim} = 1$ & $T_{\rm elim} = 2$\\
\multicolumn{1}{c}{Method} & $(n=32)$ & $(n=59)$ & $(n=71)$ & $(n=128)$ & $(n=182)$ & $(n=210)$\\
\toprule
\texttt{sel.mean} & \crdb{-0.19} & -0.44 & -0.44 & \crdb{-0.58} & \crdb{-1.33} & \crdb{-2.08} \\
\texttt{sel.min} & -0.15 & \cblb{-0.65} & \cblb{-1.19} & 2.63 & 1.47 & 0.26 \\
\texttt{ei.ord} & -0.15 & \cblb{-0.65} & \crdb{-1.40} & 2.80 & 1.76 & 1.26 \\
\texttt{ei.nom} & -0.15 & \crdb{-1.19} & \crdb{-1.40} & 2.72 & 1.28 & 0.22 \\
\texttt{sel.atm} & \crdb{-0.19} & -0.44 & \cblb{-1.19} & \cblb{-0.33} & \cblb{-0.78} & \crdb{-2.08} \\
\toprule
\multicolumn{1}{c}{} & \multicolumn{3}{c|}{\textbf{Shubert}, $p=10$} & \multicolumn{3}{c}{\textbf{Shubert}, $p=24$}\\
\multicolumn{1}{c}{} & $T_{\rm elim} = 0$ & $T_{\rm elim} = 1$ & $T_{\rm elim} = 2$ & $T_{\rm elim} = 0$ & $T_{\rm elim} = 1$ & $T_{\rm elim} = 2$\\
\multicolumn{1}{c}{Method} & $(n=48)$ & $(n=75)$ & $(n=87)$ & $(n=128)$ & $(n=182)$ & $(n=210)$\\
\toprule
\texttt{sel.mean} & -$3.0\times 10^{-6}$ & -$7.3\times 10^{-6}$ & -$1.3\times 10^{-6}$ & -$6.4\times 10^{-12}$ & -$5.7\times 10^{-11}$ & -$6.3\times 10^{-12}$\\
\texttt{sel.min} & \crdb{-}$\crdbm{1.6\times 10^{-4}}$ & \crdb{-}$\crdbm{3.0\times 10^{-4}}$ & \crdb{-}$\crdbm{6.1\times 10^{-4}}$ & \crdb{-}$\crdbm{1.4\times 10^{-8}}$ & \crdb{-}$\crdbm{4.1\times 10^{-8}}$ & \crdb{-}$\crdbm{1.2\times 10^{-7}}$ \\
\texttt{ei.ord} & \cblb{-}$\cblbm{1.5\times 10^{-4}}$ & \cblb{-}$\cblbm{2.2\times 10^{-4}}$ & -$2.5\times 10^{-4}$ & \cblb{-}$\cblbm{4.6\times 10^{-9}}$ & \cblb{-}$\cblbm{1.4\times 10^{-8}}$ & -$1.4\times 10^{-8}$\\
\texttt{ei.nom} & -$1.4\times 10^{-4}$ & -$2.0\times 10^{-4}$ & -$2.0\times 10^{-4}$ & \crdb{-}$\crdbm{1.4\times 10^{-8}}$ & \cblb{-}$\cblbm{1.4\times 10^{-8}}$ & -$1.4\times 10^{-8}$ \\
\texttt{sel.atm} & \cblb{-}$\cblbm{1.5\times 10^{-4}}$ & -$2.0\times 10^{-4}$ & \cblb{-}$\cblbm{3.0\times 10^{-4}}$ & \crdb{-}$\crdbm{1.4\times 10^{-8}}$ & \cblb{-}$\cblbm{1.4\times 10^{-8}}$ & \cblb{-}$\cblbm{4.1\times 10^{-8}}$ \\
\toprule
\end{tabular}
\normalsize
\caption{Median predicted minima $f(\hat{\bm{x}})$ over 100 trials for the DetPep10e $(p=9,18)$, Camel6 $(p=8,24)$ and Shubert $(p=10,24)$ functions. Here, $T_{\rm elim}$ denotes the number of levels eliminated for prediction at a given stage (for SEL methods), and $n$ denotes the total number of function evaluations. The best method is colored \crd{red}, and the second-best method is colored \cbl{blue}.}
\label{tbl:res}
\end{table}

\begin{table}
\centering
\footnotesize
\begin{tabular}{c c c c | c c c c c}
\toprule
\multicolumn{1}{c}{} & \multicolumn{3}{c|}{\textbf{DetPep10e}, $p=9$, noisy} & \multicolumn{5}{c}{\textbf{DetPep10e}, $p=18$, $T_{\rm elim} = 2$}\\
\multicolumn{1}{c}{} & $T_{\rm elim} = 0$ & $T_{\rm elim} = 1$ & $T_{\rm elim} = 2$ & All$\times$1 & $T_0$$\times$2 & $T_1$$\times$2 & $T_2$$\times$2 & All$\times$2\\
\multicolumn{1}{c}{Method} & $(n=32)$ & $(n=59)$ & $(n=71)$ & $(n=138)$ & $(n=202)$ & $(n=192)$ & $(n=158)$ & $(n=276)$\\
\toprule
\texttt{sel.mean} & \crdb{0.13} & \crdb{0.17} & \crdb{0.16} & \crdb{0.013} & \crdb{0.013} & \crdb{0.013} & \crdb{0.013} & \crdb{0.013}\\
\texttt{sel.min} & 3.96 & 1.25 & 0.25 & 3.195 & 3.271 & 3.324 & 3.346 & 3.249 \\
\texttt{ei.ord} & 4.09 & \cblb{0.23} & 0.19 & 2.710 & 2.659 & 2.547 & 2.661 & 2.532\\
\texttt{ei.nom} & 4.41 & 0.33 & 0.21 & 2.757 & 2.801 & 2.990 & 3.094 & 2.495\\
\texttt{sel.atm} & \cblb{0.29} & 0.24 & \cblb{0.17} & \cblb{0.038} & \cblb{0.038} & \cblb{0.036} & \cblb{0.028} & \cblb{0.022}\\
\toprule
\end{tabular}
\normalsize
\caption{\textup{(Left)} Median predicted minima $f(\hat{\bm{x}})$ over 100 trials for the DetPep10e $(p=9)$ function, with noise $\distas{i.i.d.} \mathcal{N}(0,0.5)$. \textup{(Right)} Mean predicted minima $f(\hat{\bm{x}})$ for the DetPep10e $(p=18)$ function after $T_{\rm elim} = 2$ level eliminations, under different design augmentation schemes.}
\label{tbl:resnz}
\end{table}

%Mention the sections of simulations, how it's separated. Table \ref{tbl:rec} and blah. Shubert and Shubert 24 (make EI bad).
%%Results table. Recommendations table. Roadmap for rest of section.

\subsection{Results}

Table \ref{tbl:res} reports the median of predicted minimum values from 100 simulation trials, with the best method marked in red and the second-best method in blue. Results in Table \ref{tbl:res} assume no experimental noise; the noisy setting is considered later in Table \ref{tbl:resnz} (left).

Consider first the low-dimensional simulations ($p \leq 10$ factors) on the left side of Table \ref{tbl:res}. To investigate the adaptivity of the proposed method, we focus mainly on the performance of \texttt{sel.atm} after $T_{\rm elim} = 2$ level eliminations. For DetPep10e (top-left), \texttt{sel.mean} is the best method, followed by \texttt{sel.atm} and \texttt{ei.ord}. This is expected, because the near-additive form of DetPep10e enables effective minimization using a marginal means model. Here, \texttt{sel.atm} (tuned via \texttt{tune.alpha}) learns this additive structure and adjusts to the fully model-based strategy of $\boldsymbol{\alpha} = 1$; this allows for comparable results to the gold standard of \texttt{sel.mean} after $T_{\rm elim} = 2$ eliminations. For Camel6 (middle-left), the two EI methods are the best, followed closely by \texttt{sel.atm} and \texttt{sel.min}. This is again not surprising, since the moderate interactions in Camel6 can be well modeled via a first-order autoregressive GP. Here, \texttt{sel.atm} (tuned via \texttt{tune.alpha}) detects these interactions and adjusts to a hybrid strategy with $\boldsymbol{\alpha} \in (0,1)$, which allows for comparable performance to the EI methods after $T_{\rm elim} = 2$ eliminations. For Shubert (bottom-left), the purely rank-based \texttt{sel.min} provides the best performance, followed by \texttt{sel.atm} and \texttt{ei.ord}; this is expected given the strong interactions in Shubert. Here, \texttt{sel.atm} (tuned via \texttt{tune.alpha}) detects these strong interactions and adjusts to a purely rank-based strategy of $\boldsymbol{\alpha} = 0$.

Consider next the high-dimensional simulations ($p > 10$ factors) on the right side of Table \ref{tbl:res}. One key difference here is that, compared to the earlier results, both EI techniques perform much worse than the three SEL methods. This makes intuitive sense, since GP models (which guide optimization for EI) are difficult to fit well in high-dimensions, especially with limited data \citep{Hea2018}. To contrast, the SEL methods suffer less from this ``curse-of-dimensionality'' by using \textit{marginal} information for optimization. For DetPep10e (top-right), \texttt{sel.mean} provides the best performance, followed by \texttt{sel.atm}; this shows the adaptivity of \texttt{sel.atm} in learning and exploiting additive structure for optimization. For Camel6 (middle-right), \texttt{sel.mean} is the best method, followed by \texttt{sel.atm}. This shows that, in high-dimensions with limited samples, a function with moderate interactions may be better approximated with a well-estimated main-effects model than a poorly-estimated interactions model. For Shubert (bottom-right), \texttt{sel.min} and \texttt{sel.atm} yield the best performance, which again shows the adaptivity of \texttt{sel.atm} in adjusting to a purely rank-based strategy.

%Another point of interest is the difference between the two EI methods \texttt{ei.ord} and \texttt{ei.nom}. Here, the input factors in our simulations can be viewed in one of two ways: as \textit{ordinal} discrete factors, or as \textit{nominal} discrete factors. In the case of ordinal factors, \texttt{ei.ord} (with first-order correlation \eqref{eq:ordcorr}) should be used to account for level order, whereas in the case of nominal factors, \texttt{ei.nom} (with exchangeable correlation \eqref{eq:excorr}) should be used, since no level ordering is assumed. From Table \ref{tbl:res}, we see that \texttt{ei.ord} provides better minimization to \texttt{ei.nom} in most problems (although \texttt{ei.nom} does surprisingly well in others), which suggests that incorporating order information (whenever available) does help improve the performance of EI for the current discrete problem.

Investigating next the effect of noise, Table \ref{tbl:resnz} (left) reports the median predicted minima for the DetPep10e ($p=9$) function, with observations corrupted by i.i.d. $\mathcal{N}(\mu=0,\sigma=0.5)$ noise. The two EI variants (\texttt{ei.ord}, \texttt{ei.nom}) appear to be slightly more sensitive to noise than the SEL methods (\texttt{sel.mean}, \texttt{sel.min}, \texttt{sel.atm}), in that the deterioration in performance under noise is slightly worse for the former than the latter. This robustness of SEL methods to noise can be reasoned as follows. The first method, \texttt{sel.mean}, in using marginal means for optimization and level elimination, mitigates the effect of observation noise via \textit{marginal} averaging. The second method, \texttt{sel.min}, which relies on only the \textit{ranking} of data, can be more resilient to noise compared to model-based approaches \citep{Xea2009}; this is similar to the appeal of order statistics for robust estimation \citep{Hub2011}. It is therefore not surprising that the proposed approach \texttt{sel.atm}, which incorporates both marginal estimation and ranking, also enjoys robustness to noise in optimization.

Lastly, we examine \texttt{sel.atm} under different run sizes. First, note that the earlier results employ the minimum run size needed to retain an OA structure. In practice, a run budget may be available which exceeds this minimum run size, in which case one can afford larger OAs in one or more stages of \texttt{sel.atm}. To explore this further, we take the DetPep10e ($p=18$) example with $T_{\rm elim} = 2$ level eliminations (Table \ref{tbl:res}, top-right), and compare results using the minimum run size ($n$ = 64 (initial OA) + 54 (stage 1 OA) + 20 (stage 2 OA) = 138 total runs), with three augmentation schemes which double the run size in each of the three stages ($n = 64 \times 2 + 54 + 20 = 202$, $64 + 54 \times 2 + 20 = 192$, and $64 + 54 + 20 \times 2= 158$ runs). We denote these schemes as All$\times 1$, $T_0$$\times$2, $T_1$$\times$2 and $T_2$$\times$2, respectively. Table \ref{tbl:resnz} (right) shows the mean predicted minima under these four schemes. Additional runs appear most beneficial for \texttt{sel.atm} in the second stage (i.e., $T_2$$\times$2), where the greatest objective reduction is achieved (0.028 - 0.038 = -0.010) with the smallest run size increase ($158-138 = 20$ runs). This makes sense given the near-additivity of DetPep10e: larger run sizes in early stages do not help much in estimating the large-scale additive structure, but larger run sizes in later stages are useful for identifying small-scale interactions. We therefore suggest extra runs be allocated to stages where noticeable interactions are detected. Table \ref{tbl:resnz} (right) also reports results for \texttt{sel.atm} using double the run size in \textit{all} stages (denoted as All$\times$2, with $n = (64 + 54 + 20) \times 2 = 276$ runs), which improve upon the single-stage schemes $T_0$$\times$2, $T_1$$\times$2 and $T_2$$\times$2. With a large enough budget, multi-stage augmentations can yield improved optimization over single-stage augmentations for \texttt{sel.atm}.

%From Sections \ref{sec:des} and \ref{sec:sel}, the total run size for \texttt{sel.atm} should allow for an OA design in each elimination iteration, up to a desired number of level eliminations $T_{\rm elim}$. The results in Table \ref{tbl:res} assume . but one often has a budget number of runs in practice. What to do with extra runs (allocate at beginning? allocate in middle?) ? Table \ref{tbl:resnz} (right) shows several allocations of extra runs at each elimination level for DetPep. sel.atm seems to provide the most improvement for additional runs at later elimination stages, why? and recommend? Others don't yield as dramatic an improvement, this is because the extra data allows atm to better tune its percentages for optimization (modify earlier design section to this conclusion too).

\subsection{Recommendations}

\begin{table}
\centering
\footnotesize
\begin{tabular}{c | c | c | c}
\toprule
%\multirow{2}{*}{\backslashbox{\textbf{Dim. (Type)}}{\textbf{Smoothness}}} & \multirow{2}{*}{\textit{Nearly additive}} & \textit{Non-additive}, & \textit{Non-additive},\\
% & & \textit{moderate interactions} & \textit{strong interactions}\\
%\toprule
%\multirow{2}{*}{Low-dim. (ordinal)} & \texttt{sel.mean} & \texttt{ei.ord} & \texttt{sel.min}\\
%& (\texttt{sel.atm}, \texttt{ei.ord}) & (\texttt{sel.atm}, \texttt{sel.min}) & (\texttt{sel.atm}, \texttt{ei.ord})\\
%\hline
%\multirow{2}{*}{Low-dim. (nominal)} & \texttt{sel.mean} & \texttt{ei.nom} & \texttt{sel.min} \\
%& (\texttt{sel.atm}) & (\texttt{sel.atm}, \texttt{sel.min}) & (\texttt{sel.atm})\\
%\hline
%\multirow{2}{*}{High-dim. (ordinal / nominal)} & \texttt{sel.mean} & \texttt{sel.mean} & \multirow{2}{*}{\texttt{sel.min}, \texttt{sel.atm}} \\
%& (\texttt{sel.atm}) & (\texttt{sel.atm}) & \\
\multirow{2}{*}{\backslashbox{\textbf{Smoothness}}{\textbf{Dim. (Type)}}} & \textit{Low-dim.} $p\leq10$ & \textit{Low-dim.} $p\leq10$ & \textit{High-dim.} $p > 10$\\
 & \textit{(ordinal)}& \textit{(nominal)} & \textit{(ordinal/nominal)}\\
\toprule
\multirow{2}{*}{\textit{Nearly additive}} & \texttt{sel.mean} & \texttt{sel.mean} & \texttt{sel.mean} \\
& (\texttt{sel.atm}, \texttt{ei.ord}) & (\texttt{sel.atm}) & (\texttt{sel.atm}) \\
\hline
\textit{Non-additive,} & \texttt{ei.ord} & \texttt{ei.nom} & \texttt{sel.mean} \\
\textit{moderate interactions} & (\texttt{sel.atm}, \texttt{sel.min}) & (\texttt{sel.atm}, \texttt{sel.min}) & (\texttt{sel.atm})\\
\hline
\textit{Non-additive,} & \texttt{sel.min} & \texttt{sel.min} & \texttt{sel.min} \\
\textit{strong interactions} & (\texttt{sel.atm}, \texttt{ei.ord}) & (\texttt{sel.atm}) & (\texttt{sel.atm}) \\
\toprule
\end{tabular}
\normalsize
\caption{Best (unbracketed) and second-best (bracketed) methods under different objective function smoothness, problem dimension, and type of discrete factors.}
\label{tbl:rec}
\end{table}

Summarizing these findings, Table \ref{tbl:rec} shows the best (unbracketed) and second-best (bracketed) methods under different function smoothness, problem dimension, and type of discrete factor. For \textit{low-dimensional} problems with \textit{ordinal} factors (first column), \texttt{sel.atm} and \texttt{ei.ord} both provide effective robust optimization for discrete black-box problems, where little is known on function smoothness a priori. However, for low-dimensional problems with \textit{nominal} factors (second column), \texttt{sel.atm} offers improved robust performance over existing methods. This improved robustness for \texttt{sel.atm} becomes more prominent in \textit{high-dimensional} problems (last column), where existing EI and SEL methods can yield poor solutions under certain situations. \texttt{sel.atm} is also robust to experimental noise, by employing both ranking and marginal averaging for optimization. We therefore recommend the proposed method \texttt{sel.atm} in discrete, expensive black-box problems with either (i) a large number of factors, (ii) some nominal factors, or (iii) experimental noise.

\section{Applications}

Finally, we apply \texttt{sel.atm} to two real-world applications: the first on robust parameter design of a circular piston, and the second on product family design of a thermistor network. These two applications demonstrate the lack of robustness for existing methods, and show the robust performance of \texttt{sel.atm} in practical problems.

\subsection{Robust parameter design of circular piston}
Consider first the robust parameter design for a circular piston, a key mechanical component in modern engine systems. The objectives here are two-fold: we wish to find an optimal setting of piston parts and operating conditions, which (i) achieves a target piston cycle time, but also (ii) yields as little variation as possible in cycle time when input settings deviate slightly. Objective (i) is a \textit{nominal-the-best} problem in parameter design \citep{WH2009}, where one wants to find a parameter setting with an output as close as possible to some target $C^*$. Objective (ii) is a type of robust parameter design problem, where robustness is with respect to \textit{internal noise} \citep{Tag1986} in parameters\footnote{Note that the ``robust'' parameter design in \cite{Tag1986} is different from the ``robust'' optimization tackled in this work -- the first considers the robustness of a design to internal noise in inputs, whereas the second considers the robustness of an optimization method for black-box objective functions.}. Here, internal noise can result from manufacturing tolerances on piston parts, uncontrollable variations in operating conditions, or parts degradation (Section 11.3, \citealp{WH2009}). Both objectives are important for designing an effective and reliable piston system.

Table \ref{tbl:spec} lists the $p=7$ input factors for this design problem, along with their levels and factor type. The first three factors (piston weight, piston surface area, spring choice) are controllable piston components; these factors are discrete, since components come in limited varieties by a parts supplier. The next factor permits three initial positions of the piston, which in turn affects initial gas volume $V_0$. The last three factors (pressure, ambient temperature and filling gas temperature) determine operating conditions, and are again assumed to be discrete (i.e., conditions are controllable to fixed levels). All factors are taken to be ordinal, except for initial piston position, which we assume to be nominal.

Here, we wish to design a robust parameter design $\bm{x}^*$ which yields a target cycle time of $C^* = 50$ seconds. This can be formulated as the discrete optimization problem:
\begin{equation}
\bm{x}^* = \underset{\bm{x} \in \mathcal{X}}{\text{argmin}} \; f(\bm{x}) := \underset{\bm{x} \in \mathcal{X}}{\text{argmin}} \; \max_{\bm{t} \in \mathcal{T}} |C(\bm{x} + \bm{t}) - C^*|,
\label{eq:nomrob}
\end{equation}
where $C(\bm{x})$ is the cycle time at setting $\bm{x}$, $\mathcal{T}$ is the tolerance range for internal noise, and $f(\bm{x})$ is the maximum deviation from target $C^*$ given tolerance $\mathcal{T}$, which we wish to minimize. For this problem, $\mathcal{T}$ is set as $\pm 1.5\% \Delta_l$ for each factor $l$ with design range $\Delta_l$. To mimic physical experiments, we first simulate cycle time $C(\bm{x})$ from the piston model in \cite{BS2007}, then add in i.i.d. $\mathcal{N}(\mu=0,\sigma = 0.01)$ experimental noise. Again, each evaluation of $f$ is assumed to be expensive and black-box, and the goal is to find a robust piston setting using as few runs as possible.

\begin{table}[t]
\centering
\footnotesize
\begin{tabular}{c c c c}
%\toprule
%\multicolumn{2}{c}{\textbf{Piston}}\\
\toprule
\textit{Input factors} & \textit{Levels} & \textit{Type}\\
\toprule
Piston weight & $M \in \{32.5, 37.5, 42.5, 47.5, 52.5, 57.5\}$ kg & Ordinal \\
Piston surface area & $S \in \{7.5, 12.5, 17.5\} \times 10^{-3}$ m${}^2$ & Ordinal \\
Spring choice &  $k \in \{1333,2000,2667,3333,4000,4667\}$ N/m & Ordinal \\
Initial piston position & Pos. $\{1,2,3\}$ $\Rightarrow$ $V_0 \in \{3.33,6.00,8.67\} \times 10^{-3}$ m${}^3$ & Nominal \\
Atmospheric pressure & $P_0 \in \{9.33,10.00,10.67\} \times 10^{4}$ N/m${}^2$ & Ordinal \\
Ambient temp. & $T_a \in \{291, 293, 295\}$ K & Ordinal\\
Filling gas temp. & $T_0 \in \{343.3, 350.0, 356.7\}$ K & Ordinal\\
\toprule
\end{tabular}
\caption{Input factors, level settings and factor types for the piston design problem.}
\label{tbl:spec}
\normalsize
\end{table}

We compare the performance of \texttt{sel.atm} with the two existing SEL methods \texttt{sel.mean} and \texttt{sel.min}, along with an EI method \texttt{ei.mix}, which employs a separable GP model with first-order covariance \eqref{eq:ordcorr} for ordinal factors and exchangeable covariance \eqref{eq:excorr} for nominal factors. Given a tentative budget of $n \approx 125$ runs, we run the SEL methods in two batches (i.e., with $T_{\rm elim} = 1$ level eliminations), which require $n$ = 36 (initial OA) + 100 (stage 1 OA) = 136 total runs. As in simulations, \texttt{ei.mix} is run using the same batch sample sizes. 

\begin{figure}[t]
\begin{minipage}{0.25\textwidth}
\centering
\small
\begin{tabular}{c c}
\toprule
\multicolumn{2}{c}{\textbf{Piston design}}\\
\multicolumn{2}{c}{($T_{\rm elim} = 1$, $n=136$)} \\
%\hline
Method & $f(\bm{x})$\\
\toprule
\texttt{sel.mean} & 1.13$\times$$10^{-1}$\\
\texttt{sel.min} & 1.74$\times$$10^{-2}$\\
\texttt{ei.mix} & \cblb{1.61$\times$$10^{-2}$}\\
\texttt{sel.atm} & \crdb{1.53$\times$$10^{-2}$}\\
%\hline
%0.112 & 0.017 & 0.016 & 0.015\\
\toprule
\end{tabular}
\normalsize
%\captionof{table}{Noise $\distas{i.i.d.} \mathcal{N}(0,0.1)$.}
%\label{tbl:piston}
\end{minipage}
\hspace{0.1cm}
\begin{minipage}{0.72\textwidth}
\centering
\includegraphics[width=1.02\textwidth]{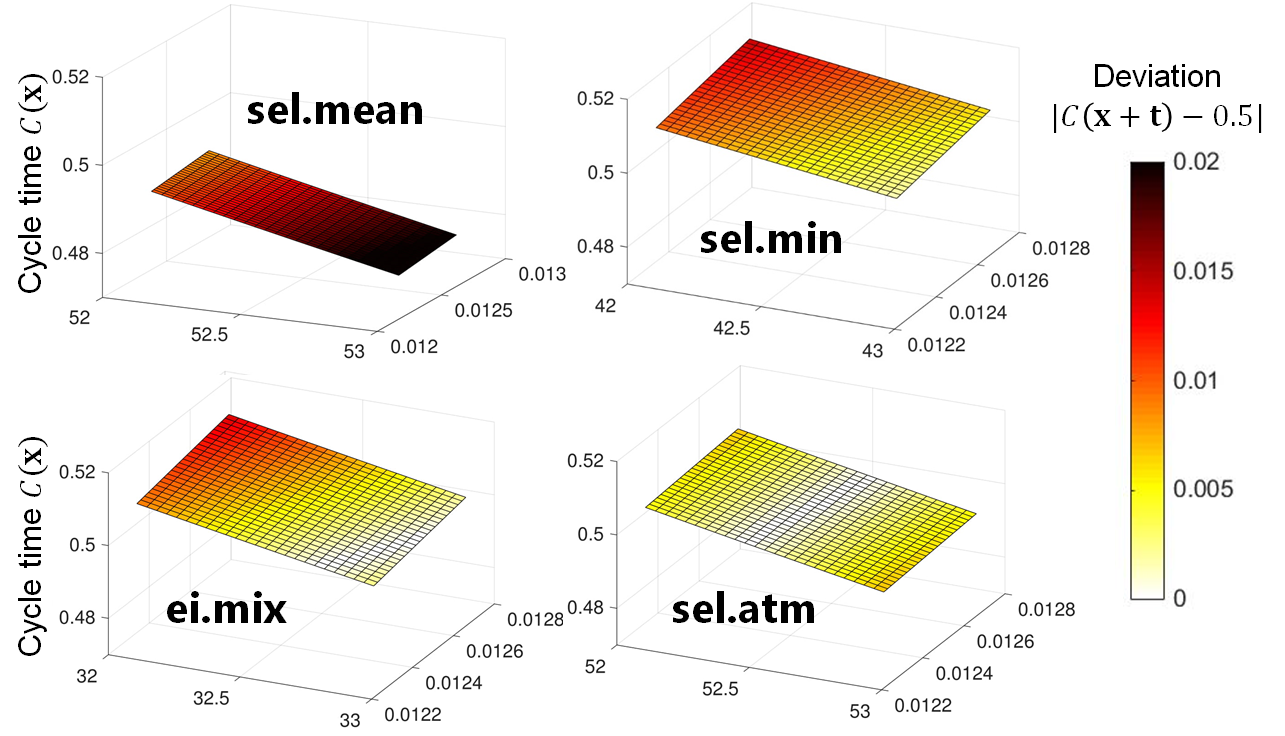}
\end{minipage}
\caption{\textup{(Left)} Median predicted minima $f(\hat{\bm{x}})$ over 100 trials for the piston design problem. $T_{\rm elim}$ denotes the number of level eliminations, and $n$ denotes the total number of runs. \textup{(Right)} Visualizing cycle time over the tolerance region of the chosen setting $\hat{\bm{x}}$, for each of the four methods.}
\label{fig:piston}
\end{figure}

Figure \ref{fig:piston} (left) reports the median predicted minima from 100 trials, with the best method in red and the second-best method in blue. Here, \texttt{sel.atm} and \texttt{sel.mean} yield the best and worst results for minimizing maximum deviation $f(\bm{x})$ from target $C^* = 0.5$, respectively. This suggests that $f$ is non-additive with moderate interactions. Such properties, however, are not known a priori for this black-box problem, and using an existing method blindly (e.g., \texttt{sel.mean}) can lead to poor piston designs, as seen here. The proposed method \texttt{sel.atm} addresses this by first detecting non-additivity from data, and exploiting it to tune a hybrid (rank- and model-based) optimization strategy. After one level elimination, this tuned strategy yield improved piston designs to all three existing methods.

To visualize robustness to internal noise, Figure \ref{fig:piston} (right) shows the cycle time $C(\bm{x})$ over the tolerance region of the chosen setting $\hat{\bm{x}}$, for the first two factors of each method. Each surface is colored by its deviation from the target, $|C(\bm{x}+\bm{t}) - C^*|$, with red and white indicating large and small deviations, respectively. Here, the surface for \texttt{sel.mean} is dark-red, which shows its piston design yields cycle times far from the target of $C^* = 0.5$. The surfaces for \texttt{sel.min} and \texttt{ei.mix} are yellow and red, meaning their designs yield times close to $C^*$ in some parts of the tolerance region, but far in other parts. The surface for \texttt{sel.atm} is white and yellow, which shows its design is the most robust to internal noise. By detecting non-additivity from data and adjusting its strategy accordingly, \texttt{sel.atm} provides effective piston settings without requiring prior knowledge on the black-box design problem.

\subsection{Product family design of thermistor network}
\begin{figure}[t]
\centering
\includegraphics[width=0.35\textwidth]{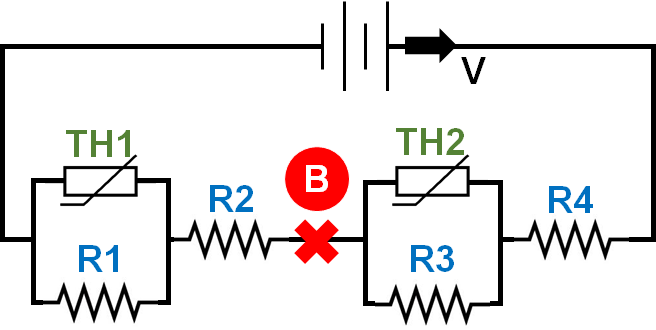}
\caption{Circuit diagram for the considered thermistor network family.}
\label{fig:cirdia}
\end{figure}

Consider next the product family design problem for a thermistor network, an important component in power supply circuits, amplifier circuits and fiber-optic communication systems. The idea behind product family design is to develop a family of products with customizable parts, then employ different combination of parts to target market-specific needs. This allows manufacturers to cut down on development and production costs, while also catering to specific demands of consumers \citep{Nea2002}. We study here the thermistor network in Figure \ref{fig:cirdia}, which is used in laser diode systems for fiber-optic communications \citep{Lyo2008}. Table \ref{tbl:circuit} lists the $p=8$ input factors considered. Six of these are customizable circuit parts: four resistors (labeled $R_1$--$R_4$), and two thermistors (resistors whose resistances depend on temperature, labeled $TH_1$ and $TH_2$). The last two factors, $V_i$ and $T_a$, control input voltage and ambient temperature. All factors are considered discrete, since resistors and thermistors come in select varieties from a parts supplier, and operating conditions are controllable to fixed levels. Here, all factors are ordinal except for $TH_1$ and $TH_2$, since both base resistance $R_b$ and temperature coefficient $\beta$ vary for the six thermistor choices.

\begin{table}[t]
%\begin{minipage}{0.2\textwidth}
%\end{minipage}
%\hspace{0.1cm}
%\begin{minipage}{0.77\textwidth}
\centering
\footnotesize
\begin{tabular}{c c c c}
\toprule
\textit{Input factors} & \textit{Levels} & \textit{Type}\\
\toprule
Resistor 1 & \multirow{4}{*}{$R_1, \cdots, R_4 \in \{500, 1000, 1500, 2000, 2500, 3000\} \; \Omega$} & Ordinal \\
Resistor 2 & & Ordinal \\
Resistor 3 & & Ordinal \\
Resistor 4 & & Ordinal \\
\hdashline
Thermistor 1 & \multicolumn{1}{l}{Choice $\{1, \cdots, 6\} \Rightarrow (R_b,\beta) \in  \{(2750,50),(3680,220),(3560,1000),$ \; \; \;} & Nominal\\
Thermistor 2 & \multicolumn{1}{r}{$(3620,2200),(3538,3300),(3930,4700)\}$} & Nominal \\
 \hdashline
Input voltage & $V_i \in \{1.05,1.10,1.15\}$ V & Ordinal \\
Ambient temp. & $T_a \in \{296.15, 298.15, 300.15\}$ K & Ordinal\\
\toprule
\end{tabular}
%\end{minipage}
\caption{Input factors, level settings and factor types for the thermistor network problem.}
\label{tbl:circuit}
\end{table}

\begin{figure}[t]
\begin{minipage}{0.25\textwidth}
\centering
\small
\begin{tabular}{c c}
\toprule
\multicolumn{2}{c}{\textbf{Thermistor network}}\\
\multicolumn{2}{c}{($T_{\rm elim} = 2$, $n=204$)} \\
\multicolumn{1}{c}{Method} & $f(\bm{x})$\\
\toprule
\texttt{sel.mean} & \crdb{4.41$\times$$10^{-3}$} \\
\texttt{sel.min} & 2.28$\times$$10^{-2}$\\
\texttt{ei.mix} & 2.25$\times$$10^{-2}$\\
\texttt{sel.atm} & \cblb{4.77$\times$$10^{-3}$}\\
%\hline
%0.112 & 0.017 & 0.016 & 0.015\\
\toprule
\end{tabular}
\normalsize
%\captionof{table}{Noise $\distas{i.i.d.} \mathcal{N}(0,0.1)$.}
%\label{tbl:piston}
\end{minipage}
\hspace{0.1cm}
\begin{minipage}{0.72\textwidth}
\centering
\includegraphics[width=0.9\textwidth]{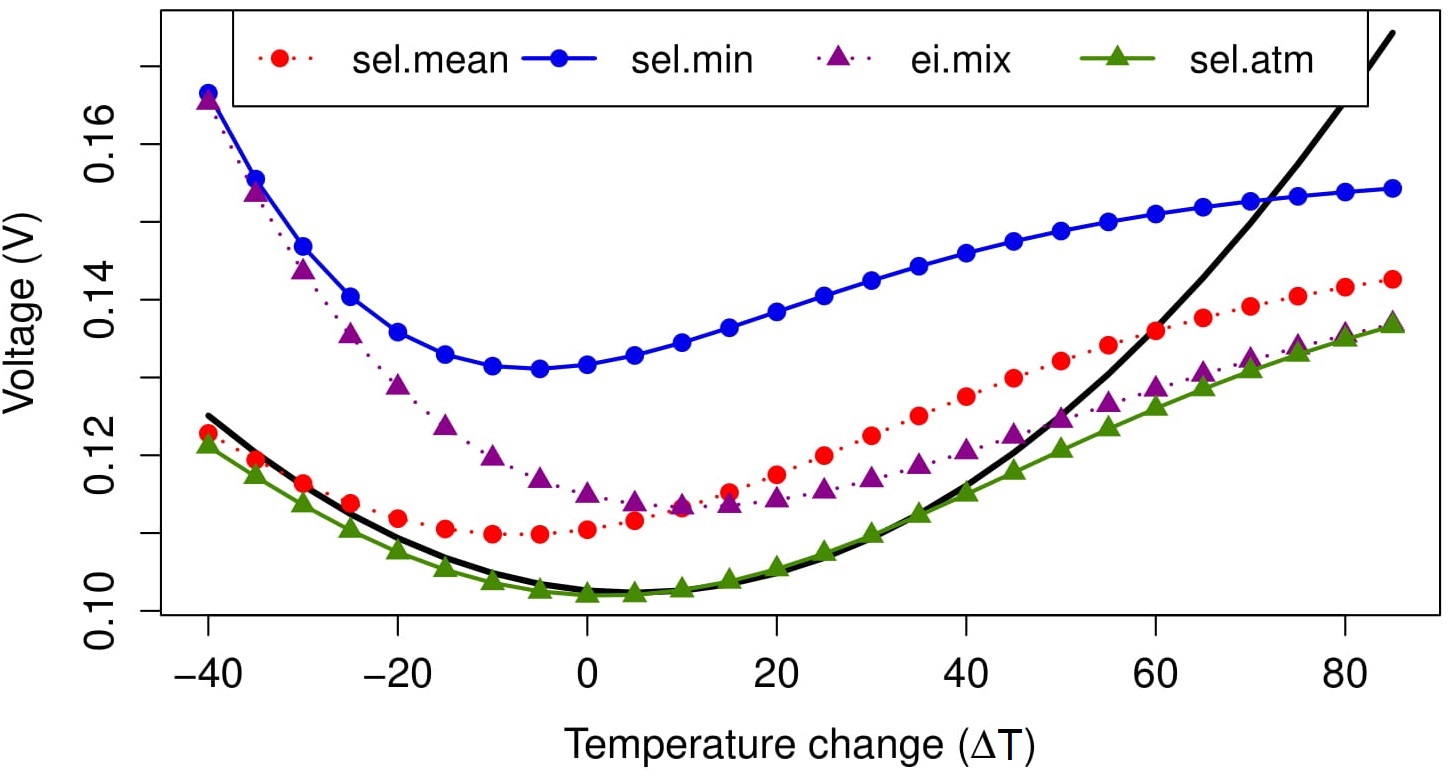}
\end{minipage}
\caption{\textup{(Left)} Median predicted minima $f(\hat{\bm{x}})$ over 100 trials for the thermistor network problem. $T_{\rm elim}$ denotes the number of level eliminations, and $n$ denotes the total number of runs. \textup{(Right)} The target $\Delta T$-$V$ curve (solid black curve), and the $\Delta T$-$V$ curves for the four methods.}
\label{fig:thermo}
\end{figure}

The goal here is to find a combination of parts and operating conditions $\bm{x}^*$ within this product family, which yields a temperature-change-voltage ($\Delta T$-$V$) curve at point B (see Figure \ref{fig:cirdia}) close to a target curve $V^*(\Delta T)$. The matching of this target curve (plotted in black, Figure \ref{fig:thermo} right) is critical to the quality of the communication device under temperature fluctuations. Using squared loss, this product design problem can be stated as:
\begin{equation}
\bm{x}^* = \underset{\bm{x} \in \mathcal{X}}{\text{argmin}} \; f(\bm{x}) := \underset{\bm{x} \in \mathcal{X}}{\text{argmin}} \int_{\Delta T} [V^*(\Delta T) - V(\Delta T; \bm{x})]^2 \; d\Delta T,
\label{eq:nomrob}
\end{equation}
where $V(\Delta T; \bm{x})$ is the $\Delta T$-$V$ curve under setting $\bm{x}$. We mimic the physical experiment for $V(\Delta T; \bm{x})$ via a simulation module in \textsf{MATLAB} \citep{Del2016}, then add in i.i.d. $\mathcal{N}(\mu=0, \sigma=0.005)$ experimental noise. \texttt{sel.atm} is compared with the SEL methods \texttt{sel.mean} and \texttt{sel.min}, as well as the EI method \texttt{ei.mix}. With a tentative budget of $n \approx 200$ runs, the SEL methods are run in three batches (i.e., $T_{\rm elim} = 2$ eliminations), which require $n$ = 36 (initial OA) + 100 (stage 1 OA) + 16 (stage 2 OA) = 204 total runs; \texttt{ei.mix} is also run using the same batch sizes. Each product design test is assumed to be expensive and black-box, and we would like to find a good design using as few runs as possible.

Figure \ref{fig:thermo} (left) reports the median predicted minima from 100 trials, with the best method in red and the second-best method in blue. Compared to the earlier piston application, the performance of the three existing methods has now \textit{reversed}: \texttt{sel.mean} yields the best results, with \texttt{ei.mix} and \texttt{sel.min} performing poorly. This suggests that $f$ is nearly additive with small interactions. However, this near-additivity of $f$ is unknown prior to experimentation, and using an existing method blindly (e.g., \texttt{ei.mix} or \texttt{sel.min}) can lead to poor product designs. The proposed method \texttt{sel.atm} addresses this by first detecting the additivity of $f$ from data, then adjusting its optimization strategy to exploit this structure. After two level eliminations, this adaptive strategy allows \texttt{sel.atm} to provide similar results to the gold standard of \texttt{sel.mean}.

Figure \ref{fig:thermo} (right) shows the $\Delta T$-$V$ curves for the product designs from the four methods, along with the target curve $V^*(\Delta T)$ in black. The $\Delta T$-$V$ curves for \texttt{sel.min} and \texttt{ei.mix} both fit the black curve poorly, meaning the product designs from these two methods do not meet target specifications. The curves for \texttt{sel.mean} and \texttt{sel.atm} provide improved fits, meaning their product designs better match target requirements (under squared loss). While there are some discrepancies to the target $V^*(\Delta T)$ for the latter two methods, this may be due to the limited specification in Table \ref{tbl:circuit}. By detecting additivity from data and exploiting this additivity in optimization, \texttt{sel.atm} provides effective product design optimization with limited data, without prior knowledge on the black-box problem. 

These two applications nicely demonstrate the lack of robustness of existing methods: \texttt{ei.mix} performs well for the first problem but poorly for the second, whereas the opposite is true for \texttt{sel.mean}. The proposed method \texttt{sel.atm}, by adaptively tuning the trade-off between rank- and model-based optimization, provides effective robust optimization for expensive black-box problems.

\section{Conclusion}

We propose in this paper a new method, Analysis-of-marginal-Tail-Means (ATM), which provides effective robust optimization for a broad range of discrete black-box problems with limited data. ATM addresses a key weakness in existing methods: these methods perform well under certain model assumptions, but yield poor performance when such assumptions (which are difficult to verify in black-box problems) are violated. The key novelty in ATM is the use of marginal tail means for optimization, which trades-off between rank-based and model-based methods. To tune this trade-off, we use a technique called \texttt{tune.alpha}, which first identifies important main effects and interactions from data, then finds a good compromise between rank- and model-based optimization to best exploit additive structure. We then introduce a batch-sequential scheme for ATM, called \texttt{sel.atm}, which uses level eliminations to quickly target regions of interest. Finally, we show in simulations and applications the lack of robustness for existing methods, and how the proposed method \texttt{sel.atm} offers more effective and robust optimization in real-world engineering problems.

%Combining everything together in a novel level elimination algorithm called \texttt{sel.atm}, we demonstrate using simulations and real-world applications how \texttt{sel.atm} offers more effective and robust performance over existing parameter optimization methods.

Looking ahead, there are several directions to pursue next. Given the effectiveness of ATM in adaptively exploiting marginal structure, we are exploring other hybrid (rank- and model-based) methods which employ different modeling approaches for optimization (e.g., EI). There is also recent work on a new type of main effect, called a \textit{conditional main effect} \citep{Wu2015, SW2017, MW2017}, which detects conditional additive structure; it would be interesting to see if incorporating such effects can improve ATM.

%direction is to further improve the proposed  algorithm \texttt{sel.atm} using recent developments on the sequential level elimination method, including the genetic algorithm and Gaussian process modifications in \cite{Mea2006} and \cite{Mea2009}, respectively. Another potentially fruitful direction is to apply the methods in this paper for warm-starting and accelerating large-scale integer programming problems. Lastly, the conditional nature of the MCR condition in \eqref{eq:fnform} appears to have some connections to recent work on conditional main effects \citep{Wu2015, SW2017, MW2017}, and it would be interesting to see whether an alternate modeling approach from this body of work can yield improved optimization performance for \texttt{sel.atm}.

\spacingset{1.3}
\small
\bibliography{references}

\begin{thebibliography}{}

\bibitem[Ali et~al., 2005]{Aea2005}
Ali, M.~M., Khompatraporn, C., and Zabinsky, Z.~B. (2005).
\newblock A numerical evaluation of several stochastic algorithms on selected
  continuous global optimization test problems.
\newblock {\em Journal of Global Optimization}, 31(4):635--672.

\bibitem[Baker and Filbeck, 2014]{BF2014}
Baker, H.~K. and Filbeck, G. (2014).
\newblock {\em Investment Risk Management}.
\newblock Oxford University Press.

\bibitem[Barone~Adesi, 2016]{Bar2016}
Barone~Adesi, G. (2016).
\newblock Va{R} and {CVaR} implied in option prices.
\newblock {\em Journal of Risk and Financial Management}, 9(1):2.

\bibitem[Ben-Ari and Steinberg, 2007]{BS2007}
Ben-Ari, E.~N. and Steinberg, D.~M. (2007).
\newblock Modeling data from computer experiments: an empirical comparison of
  kriging with {MARS} and projection pursuit regression.
\newblock {\em Quality Engineering}, 19(4):327--338.

\bibitem[Bien et~al., 2013]{Bea2013}
Bien, J., Taylor, J., and Tibshirani, R. (2013).
\newblock A lasso for hierarchical interactions.
\newblock {\em The Annals of Statistics}, 41(3):1111--1141.

\bibitem[DeLand, 2016]{Del2016}
DeLand, S. (2016).
\newblock Optimal component selection using the mixed-integer genetic
  algorithm.
\newblock
  https://www.mathworks.com/matlabcentral/fileexchange/35810-optimal-component-selection-using-the-mixed-integer-genetic-algorithm.

\bibitem[Dette and Pepelyshev, 2010]{DP2010}
Dette, H. and Pepelyshev, A. (2010).
\newblock Generalized {L}atin hypercube design for computer experiments.
\newblock {\em Technometrics}, 52(4):421--429.

\bibitem[Dimopoulos and Zalzala, 2000]{DZ2000}
Dimopoulos, C. and Zalzala, A.~M. (2000).
\newblock Recent developments in evolutionary computation for manufacturing
  optimization: problems, solutions, and comparisons.
\newblock {\em IEEE Transactions on Evolutionary Computation}, 4(2):93--113.

\bibitem[Friedman et~al., 1983]{Fea1983}
Friedman, J.~H., Grosse, E., and Stuetzle, W. (1983).
\newblock Multidimensional additive spline approximation.
\newblock {\em SIAM Journal on Scientific and Statistical Computing},
  4(2):291--301.

\bibitem[Groemping, 2017]{Gro2017}
Groemping, U. (2017).
\newblock {\em DoE.base: Full Factorials, Orthogonal Arrays and Base Utilities
  for DoE Packages}.
\newblock R package version 0.30.

\bibitem[Harari et~al., 2018]{Hea2018}
Harari, O., Bingham, D., Dean, A., and Higdon, D. (2018).
\newblock Computer experiments: Prediction accuracy, sample size and model
  complexity revisited.
\newblock {\em Statistica Sinica}, 28:899--919.

\bibitem[Hardy, 2006]{Har2006}
Hardy, M.~R. (2006).
\newblock An introduction to risk measures for actuarial applications.
\newblock {\em SOA Syllabus Study Note}.

\bibitem[Hedayat et~al., 2012]{Hea2012}
Hedayat, A.~S., Sloane, N. J.~A., and Stufken, J. (2012).
\newblock {\em Orthogonal Arrays: Theory and Applications}.
\newblock Springer Science \& Business Media.

\bibitem[Huber, 1974]{Hub2011}
Huber, P.~J. (1974).
\newblock {\em Robust Statistics}.
\newblock Wiley, New York.

\bibitem[Jamil and Yang, 2013]{JY2013}
Jamil, M. and Yang, X.-S. (2013).
\newblock A literature survey of benchmark functions for global optimisation
  problems.
\newblock {\em International Journal of Mathematical Modelling and Numerical
  Optimisation}, 4(2):150--194.

\bibitem[Johnson et~al., 2008]{Jea2008}
Johnson, K., Mandal, A., and Ding, T. (2008).
\newblock Software for implementing the sequential elimination of level
  combinations algorithm.
\newblock {\em Journal of Statistical Software}, 25(6):1--13.

\bibitem[Johnson et~al., 1990]{Jea1990}
Johnson, M.~E., Moore, L.~M., and Ylvisaker, D. (1990).
\newblock Minimax and maximin distance designs.
\newblock {\em Journal of Statistical Planning and Inference}, 26(2):131--148.

\bibitem[Jones et~al., 1998]{Jea1998}
Jones, D.~R., Schonlau, M., and Welch, W.~J. (1998).
\newblock Efficient global optimization of expensive black-box functions.
\newblock {\em Journal of Global Optimization}, 13(4):455--492.

\bibitem[Joseph et~al., 2015]{Jea2015}
Joseph, V.~R., Dasgupta, T., Tuo, R., and Wu, C. F.~J. (2015).
\newblock Sequential exploration of complex surfaces using minimum energy
  designs.
\newblock {\em Technometrics}, 57(1):64--74.

\bibitem[Lindroth and Patriksson, 2011]{LP2011}
Lindroth, P. and Patriksson, M. (2011).
\newblock Pure categorical optimization: A global descent approach.
\newblock Technical report, Division of Mathematics, University of Gothenburg.

\bibitem[Lyon, 2008]{Lyo2008}
Lyon, C. (2008).
\newblock Genetic algorithm solves thermistor-network component values.
\newblock
  https://www.edn.com/design/analog/4326942/Genetic-algorithm-solves-thermistor-network-component-values.

\bibitem[Mak et~al., 2018]{Mea2017}
Mak, S., Sung, C.~L., Wang, X., Yeh, S.~T., Chang, Y.~H., Joseph, V.~R., Yang,
  V., and Wu, C. F.~J. (2018).
\newblock An efficient surrogate model for emulation and physics extraction of
  large eddy simulations.
\newblock {\em Journal of the American Statistical Association}.
\newblock To appear.

\bibitem[Mak and Wu, 2018]{MW2017}
Mak, S. and Wu, C. F.~J. (2018).
\newblock cmenet: a new method for bi-level variable selection of conditional
  main effects.
\newblock {\em Journal of the American Statistical Association}.
\newblock To appear.

\bibitem[Mandal et~al., 2009]{Mea2009}
Mandal, A., Ranjan, P., and Wu, C. F.~J. (2009).
\newblock $\mathcal{G}$-{SELC}: Optimization by sequential elimination of level
  combinations using genetic algorithms and {G}aussian processes.
\newblock {\em The Annals of Applied Statistics}, 3(1):398--421.

\bibitem[Mandal et~al., 2006]{Mea2006}
Mandal, A., Wu, C. F.~J., and Johnson, K. (2006).
\newblock {SELC}: Sequential elimination of level combinations by means of
  modified genetic algorithms.
\newblock {\em Technometrics}, 48(2):273--283.

\bibitem[McKay et~al., 1979]{Mea1979}
McKay, M.~D., Beckman, R.~J., and Conover, W.~J. (1979).
\newblock Comparison of three methods for selecting values of input variables
  in the analysis of output from a computer code.
\newblock {\em Technometrics}, 21(2):239--245.

\bibitem[Nayak et~al., 2002]{Nea2002}
Nayak, R.~U., Chen, W., and Simpson, T.~W. (2002).
\newblock A variation-based method for product family design.
\newblock {\em Engineering Optimization}, 34(1):65--81.

\bibitem[Ranjan et~al., 2008]{Rea2008}
Ranjan, P., Bingham, D., and Michailidis, G. (2008).
\newblock Sequential experiment design for contour estimation from complex
  computer codes.
\newblock {\em Technometrics}, 50(4):527--541.

\bibitem[Simpson, 2004]{Sim2004}
Simpson, T.~W. (2004).
\newblock Product platform design and customization: Status and promise.
\newblock {\em Artificial Intelligence for Engineering Design, Analysis and
  Manufacturing}, 18(1):3--20.

\bibitem[Stein, 2012]{Ste2012}
Stein, M.~L. (2012).
\newblock {\em Interpolation of Spatial Data: Some Theory for Kriging}.
\newblock Springer Science \& Business Media.

\bibitem[Su and Wu, 2017]{SW2017}
Su, H. and Wu, C. F.~J. (2017).
\newblock {CME} analysis: a new method for unraveling aliased effects in
  two-level fractional factorial experiments.
\newblock {\em Journal of Quality Technology}, 49(1):1--10.

\bibitem[Taguchi, 1986]{Tag1986}
Taguchi, G. (1986).
\newblock {\em Introduction to Quality Engineering: Designing Quality into
  Products and Processes}.
\newblock Quality Resources.

\bibitem[Williams et~al., 2000]{Wea2000}
Williams, B.~J., Santner, T.~J., and Notz, W.~I. (2000).
\newblock Sequential design of computer experiments to minimize integrated
  response functions.
\newblock {\em Statistica Sinica}, 10:1133--1152.

\bibitem[Wu, 2015]{Wu2015}
Wu, C. F.~J. (2015).
\newblock Post-{F}isherian experimentation: from physical to virtual.
\newblock {\em Journal of the American Statistical Association},
  110(510):612--620.

\bibitem[Wu and Hamada, 2009]{WH2009}
Wu, C. F.~J. and Hamada, M.~S. (2009).
\newblock {\em Experiments: {P}lanning, {A}nalysis, and {O}ptimization}.
\newblock John Wiley \& Sons.

\bibitem[Wu et~al., 1987]{Wea1987}
Wu, C. F.~J., Mao, S.~S., and Ma, F.~S. (1987).
\newblock An investigation of {OA}-based methods for parameter design
  optimization.
\newblock Technical Report No. 24, Center for Quality and Productivity
  Improvement, University of Wisconsin-Madison.

\bibitem[Wu et~al., 1990]{Wea1990}
Wu, C. F.~J., Mao, S.~S., and Ma, F.~S. (1990).
\newblock {SEL}: {A} search method based on orthogonal arrays.
\newblock In {\em Statistical Design and Analysis of Industrial Experiments (S.
  Ghosh, ed.)}, pages 279--310. Marcel Dekker.

\bibitem[Xu et~al., 2009]{Xea2009}
Xu, S., Adiga, N., Ba, S., Dasgupta, T., Wu, C. F.~J., and Wang, Z.~L. (2009).
\newblock Optimizing and improving the growth quality of {Z}n{O} nanowire
  arrays guided by statistical design of experiments.
\newblock {\em ACS Nano}, 3(7):1803--1812.

\bibitem[Yeh et~al., 2018]{Yea2018}
Yeh, S.-T., Wang, X., Sung, C.-L., Mak, S., Chang, Y.-H., Zhang, L., Wu, C.
  F.~J., and Yang, V. (2018).
\newblock Common proper orthogonal decomposition-based spatiotemporal emulator
  for design exploration.
\newblock {\em AIAA Journal}, 56(6):2429--2442.

\end{thebibliography}
\normalsize

\pagebreak

\section*{Appendix}
\subsection*{Proof of Proposition \ref{thm:amc}}
\begin{proof}
Choose an arbitrary setting $\bm{x} \in \mathcal{X}$. Note that:
\begin{align}
f(\bm{x}) &= f(x_1,x_2, \cdots, x_p) \notag\\
&\geq f(\hat{x}_1, x_2, \cdots, x_p) \tag{by the MC condition in \eqref{eq:fnform}}\\
&\geq f(\hat{x}_1, \hat{x}_2, \cdots, x_p) \tag{again, by \eqref{eq:fnform}}\\
& \quad \vdots \notag \\
&\geq f(\hat{x}_1, \hat{x}_2, \cdots, \hat{x}_p) \notag\\
& = f(\hat{\bm{x}}_{\rm AM}). \notag
\end{align}
Hence, $f(\bm{x}) \geq f(\hat{\bm{x}}_{\rm AM})$ for any $\bm{x} \in \mathcal{X}$, which means $\hat{\bm{x}}_{\rm AM}$ must be an optimal solution.
\end{proof}

%\section{Proof of Proposition \ref{thm:pw}}
%\begin{proof}
%This can easily be shown using first principles. Fix a particular factor $l = 1, \cdots, p$, and let $\mathcal{M}_l$ be the marginal minimum statistic $\mathcal{M}_{\rm min}$. Note that:
%\[\hat{x}_l = \underset{x_l \in [N_l]}{\text{argmin}} \; \hat{m}_l(x_l) = \underset{x_l \in [N_l]}{\text{argmin}} \; \left\{ \min_{y \in \mathcal{F}_l(x_l)} y \right\} = x_l^*,\]
%where the last step follows because the double minimum over level $x_l \in [N_l]$ and marginal slice $\mathcal{F}_l(x_l)$ returns the smallest function value of $f$ on $\mathcal{X}$. Repeating this argument for every factor $l \in 1, \cdots, p$ proves the statement. 
%\end{proof}

\end{document}